\documentclass[aps,eqsecnum,preprint,floats,epsf,epsfig,nofootinbib,showpacs]{revtex4}
\usepackage{graphicx}
\def\be{\begin{eqnarray}}
\def\en{\end{eqnarray}}
\def\non{\nonumber}
\def\la{\langle}
\def\ra{\rangle}

\def\bi{\bibitem}

\begin{document}

\title{\Large \bf Study of light-cone distribution amplitudes
for $p$-wave heavy mesons
 }

\author{ \bf  Chien-Wen Hwang\footnote{
t2732@nknucc.nknu.edu.tw}}

\affiliation{\centerline{Department of Physics, National Kaohsiung Normal University,} \\
\centerline{Kaohsiung, Taiwan 824, Republic of China}
 }


\begin{abstract}
In this paper, a study of light-cone distribution amplitudes
for $p$-wave heavy mesons is presented in both general and heavy
quark frameworks. Within the light-front approach, the leading twist
light-cone distribution amplitudes, $\phi_M(u)$ and their relevant
decay constants of heavy scalar, axial-vector and tensor mesons, $f_M$,
are formulated. The relations of some decay constants can be simplified when the
heavy quark limit is taken into account. After fixing the
parameters which appear in a Gaussian wave
function, the corresponding decay constants are calculated and
compared with those of other theoretical approaches. The curves and
the first six $\xi$ moments of $\phi_M(u)$ are plotted and
estimated. These results all endorse the requirements of heavy quark symmetry.
\end{abstract}
\maketitle %

\section{Introduction}
Light-cone distribution amplitudes (LCDAs) of hadrons are key
ingredients in the description of various exclusive processes of
quantum chromodynamics (QCD), and their role can be analogous to
those of parton distributions in inclusive processes. In terms of
Bethe-Salpeter wave functions $\varphi(u_i,k_{i\perp})$, LCDAs
$\phi(u_i)$ are defined by retaining the momentum fractions $u_i$
and integrating out the transverse momenta $k_{i\perp}$ \cite{LB}.
They provide essential information on the nonperturbative structure
of the hadron for QCD treatment of exclusive reactions.
Specifically, the leading twist LCDAs describe the probability
amplitudes to find the hadron in a Fock state with the minimum
number of constituents. In the
literature, there have been many nonperturbative approaches to
estimate LCDAs, such as the QCD sum rules
\cite{CZ,Bakulev,Ball1,Yang1,Bragutanew}, lattice calculation
\cite{Ali,Braun}, chiral quark model from the instanton vacuum
\cite{Petrov,Nam}, Nabmbu-Jona-Lasinio model
\cite{Arriola,Praszalowicz}, and the light-front quark model
\cite{Ji1,me,Ji2}. These studies have dealt with LCDAs of
pseudoscalar
\cite{Bakulev,Braun,Petrov,Nam,Arriola,Praszalowicz,Ji1,me}, vector
\cite{Ball1,Ali,Ji1,me}, axial-vector \cite{Yang1,Bragutanew,Ji2},
and tensor \cite{Bragutanew} mesons.

The fact that $B$-physics
exclusive processes are under investigation in BABAR, Belle, and LHC
experiments also urges the detailed study of hadronic LCDAs.
Recently, many $p$-wave heavy mesons were observed and confirmed.
They include $D^{*0}_0$, $D_1$, $D^{*}_2$, $D^{\pm}_{s1}$, $D^{*}_{s2}$
\cite{Babar0,Babar1,ZEUS,D0,LHCb} and $B_1^0$, $B^{*0}_2$, $B^0_{s1}$,
$B^{*0}_{s2}$ \cite{CDF0,CDF1,D0a}. The present paper is devoted to
the study of leading twist LCDAs of $p$-wave heavy mesons which
include the scalar, axial-vector, and tensor mesons. 
We hope a thorough understanding of their properties, such as LCDAs
which are universal nonperturbative objects, will be of great
benefit when analyzing the hard exclusive processes with heavy meson
production.


In the past decade, the most significant progress made in the QCD
description of hadronic physics was, perhaps, in the avenue of heavy
quark dynamics. The analysis of heavy hadron structures has been
tremendously simplified by the heavy quark symmetry (HQS) proposed
by Isgur and Wise \cite{IW1,IW2}, and the heavy quark effective 
theory (HQET) developed from QCD in terms of $1/m_Q$ expansion 
\cite{Georgi,EH1,EH2}. HQET has provided
a systematic framework for studying symmetry breaking $1/m_Q$
corrections (for a review, see Ref. \cite{Neubert}). Moreover, in terms
of heavy quark expansion, HQET offered a new framework for the
systematic study of the inclusive decays of heavy mesons
\cite{CGG,Bigi,MW1,Mannel}. However, the general properties of heavy
hadrons, namely, their decay constants, transition form factors,
structure functions, etc, are still incalculable within QCD, even in
the infinite quark-mass limit with the utilization of HQS and HQET.
Hence, although HQS and HQET have simplified heavy quark dynamics, a
complete first-principles QCD description of heavy hadrons is still 
lacking due to the unknown nonperturbative QCD dynamics.

In this study, the $p$-wave heavy meson is
explored in a light-front quark model (LFQM) with both general and
heavy quark frameworks. LFQM is a
promising analytic method for solving the nonperturbative problems
of hadron physics \cite{BPP}, as well as offering much insights into
the internal structures of bound states. The basic ingredient in
LFQM is the relativistic hadron wave function which generalizes
distribution amplitudes by including transverse momentum
distributions; it contains all the information of a hadron from its
constituents. The hadronic quantities are represented by the overlap
of wave functions and can be derived in principle. The light-front
wave function is manifestly a Lorentz-invariant, expressed in terms
of internal momentum fraction variables which are independent of the
total hadron momentum. Moreover, the fully relativistic treatment of
quark spins and center-of-mass motion can be carried out using the
so-called Melosh rotation \cite{LFQM}. This treatment has been
successfully applied to calculate phenomenologically many important
meson decay constants and hadronic form factors \cite{Jaus1, CCH1,
Jaus2, CCH2, Hwang}. Therefore, the main purpose of this study is
the calculation of the leading twist LCDAs of $p$-wave heavy
mesons within LFQM.

The remainder of this paper is organized as follows. In Sec. II,
the leading twist LCDAs of $p$-wave meson states are shown in cases
of vector and tensor currents with general and heavy quark frames.
In Sec. III, the formulism of LFQM is reviewed briefly; then, the
leading twist LCDAs are extracted within the LFQM.
In Sec. IV, numerical results of the decay constants and LCDAs are recorded.
The $\xi$ moments of these LCDAs are also calculated and presented.
Finally, conclusions are given in Sec. V.

\section{Leading twist LCDAs of $p$-wave mesons}
\subsection{General Framework}
Amplitudes of hard processes involving $p$-wave mesons can be
described by the matrix elements of gauge-invariant nonlocal
operators, which are sandwiched between the vacuum and the meson
states:
 \be
 \langle 0 |\bar q (x) \Gamma [x,-x] q(-x) | H(P,\epsilon)\rangle,
 \label{nonlocal}
 \en
where $P$ is the meson momentum, $\epsilon$ is the polarization
vector or tensor ($\epsilon$ does not exist in the case
of the scalar meson), $\Gamma$ is a generic notation for the Dirac
matrix structure, and the path-ordered gauge factor is
 \be
 [x,y]={\textrm{P exp}}\left[ig_s\int^1_0 dt(x-y)_\mu
 A^\mu(tx+(1-t)y)\right].
 \en
This factor is equal to unity in the light-cone gauge which is
equivalent to the fixed-point gauge, $(x-y)_\mu A^\mu (x-y)=0$, as
the quark-antiquark pair is at the lightlike separation
\cite{Yang2}. For simplicity, the gauge factor will not be shown
below.

The asymptotic expansion of exclusive amplitudes, in powers with large
momentum transfer, is governed by the expanding amplitude Eq.
(\ref{nonlocal}), shown in powers of deviation from the light-cone
$x^2=0$. There are two lightlike vectors, $p$ and $z$, which can be
introduced by
 \be
 p^2=0,~~~~~~z^2=0,
 \en
such that $p \to P$ in the limit $M_H^2 \to 0$ and $z \to x$ for
$x^2 = 0$. From this, it follows that \cite{Ball1}
 \be
 z^\mu &=& x^\mu - P^\mu \frac{1}{M_H^2} \left[Px-\sqrt{(Px)^2-x^2
 M_H^2}\right] \non \\
 &=& x^\mu - P^\mu \frac{x^2}{2 P z}+ O(x^4), \non \\
  p^\mu &=& P^\mu -z^\mu \frac{M^2_H}{2 P z},\label{Pp}
 \en
where $P x \equiv P \cdot x$ and $P z=p z=\sqrt{(P x)^2-x^2 M_H^2}$.
In addition, if it is assumed that the meson moves in a positive
$\hat{e}_3$ direction, then $p^+$ and $z^-$ are the only nonzero
components of $p$ and $z$, respectively, in an infinite momentum
frame. For the axial-vector meson, the polarization vector
$\epsilon^\mu$ is decomposed into longitudinal and transverse
projections as
 \be
 \epsilon^\mu_{\|}=\frac{\epsilon z}{p z}\left(p^\mu-z^\mu\frac{M^2_H}{2p z}\right),
 ~~~\epsilon^\mu_{\perp}=\epsilon^\mu-\epsilon^\mu_{\|}, \label{epsilon}
 \en
respectively. For the tensor meson, the polarization tensor is
 \be
 \epsilon^{\mu\nu}(m)=\langle 11;m'm''|11;2m \rangle \epsilon^\mu (m')
 \epsilon^\nu (m''),
 \en
($m$ is the magnetic quantum number) or
 \be
 \epsilon^{\mu\nu}_{\pm 2}&=&\epsilon^{\mu}_{\pm 1}\epsilon^{\nu}_{\pm 1},\\
 \epsilon^{\mu\nu}_{\pm 1}&=&\sqrt{\frac{1}{2}}\big[\epsilon^{\mu}_{\pm 1}\epsilon^{\nu}_0+\epsilon^{\mu}_0\epsilon^{\nu}_{\pm 1}\big],\\
 \epsilon^{\mu\nu}_{0}&=&\sqrt{\frac{1}{6}}\big[\epsilon^{\mu}_{+ 1}\epsilon^{\nu}_{-1}+\epsilon^{\mu}_{-1}\epsilon^{\nu}_{+1}\big]
 +\sqrt{\frac{2}{3}}\epsilon^{\mu}_{0}\epsilon^{\nu}_{0},
 \en
and $\epsilon_{\mu\bullet}
(\equiv \epsilon^{\mu\nu} z_\nu)$ can also be decomposed into
longitudinal and transverse projections as
 \be
 \epsilon^{\mu\bullet}_\|=\frac{\epsilon^{\bullet\bullet}}{p z}\left(p^\mu-z^\mu\frac{M^2_H}{2p z}\right),
 ~~~\epsilon^{\mu\bullet}_\perp=\epsilon^{\mu\bullet}-\epsilon^{\mu\bullet}_\|.
 \en

LCDAs are defined in terms of the matrix element of a nonlocal operator in
Eq. (\ref{nonlocal}). For scalar $(S)$, axial vector $(A)$, and
tensor $(T)$ mesons, the leading twist LCDAs can be defined as
 \be
 \langle 0|\bar q (z) \gamma^\mu q (-z)|S(P)\rangle &=& f_S \int^1_0 du~e^{i\xi p z}\left[p^\mu
  \phi_S(u)+z^\mu \frac{M_S^2}{2p z}g_S(u)\right], \label{S}\\
 \langle 0|\bar q (z) \gamma^\mu \gamma_5 q (-z)|A(P,\epsilon_{\lambda=0})\rangle
 &=&if_A M_A \int^1_0 du~e^{i\xi p z}\Big\{p^\mu \frac{\epsilon z}{p z}
 \phi_{A\|}
 (u) +\epsilon^\mu_{\perp}g_{A\perp}(u)\non \\
 &&~~~~~~~~~~~~~~~~~~~~~~- z^\mu \frac{\epsilon z}{2 (p z)^2}M^2_A
 g_{A3}(u)\Big\}, \label{AL}
 \en
 \be
 &&\langle 0|\bar q (z) \sigma^{\mu\nu} \gamma_5 q (-z)|A(P,\epsilon_{\lambda=\pm 1})\rangle
 =f^\perp_A \int^1_0 du~e^{i\xi p z}\Big\{(\epsilon^\mu_{\perp} p^\nu-\epsilon^\nu_{\perp} p^\mu)
 \phi_{A\perp} (u)\non \\
 &&\qquad\qquad\qquad\qquad\qquad\qquad+(p^\mu z^\nu-p^\nu z^\mu)\frac{M^2_A \epsilon z}{(p z)^2}h_{A\|}(u)
 +(\epsilon^\mu_{\perp} z^\nu-\epsilon^\nu_{\perp} z^\mu) \frac{M^2_A}{2 p z}
 h_{A3}(u)\Big\},\non \\ \label{AT}\\
 &&\langle 0|\bar q (z) \gamma^\mu q (-z)|T(P,\epsilon_{\lambda=0})\rangle
 =f_T M_T^2 \int^1_0 du~e^{i\xi p z}\Big\{p^\mu \frac{\epsilon^{\bullet\bullet}}{(p z)^2}
 \phi_{T\|}(u) +\frac{\epsilon^{\mu\bullet}_\perp}{p z}g_{T\perp}(u)\non \\
 &&\qquad\qquad\qquad\qquad\qquad\qquad- z^\mu \frac{\epsilon^{\bullet\bullet}}
 {2 (p z)^3}M^2_T g_{T3}(u)\Big\},\label{TL} \\
 &&\langle 0|\bar q (z) \sigma^{\mu\nu} q (-z)|T(P,\epsilon_{\lambda=\pm 1})\rangle
 =if^\perp_T M_T \int^1_0 du~e^{i\xi p
 z}\Big\{\frac{(\epsilon^{\mu\bullet}_\perp
 p^\nu-\epsilon^{\bullet\nu}_\perp
 p^\mu)}{p z}
 \phi_{T\perp} (u)\non \\
 &&\qquad\qquad\qquad\qquad\qquad+(p^\mu z^\nu-p^\nu z^\mu)\frac{M^2_T
 \epsilon^{\bullet\bullet}}{(p z)^3}h_{T\|}(u)
 +(\epsilon^{\mu\bullet}_\perp z^\nu-\epsilon^{\bullet\nu}_\perp z^\mu) \frac{M^2_T}{2 (p z)^2}
 h_{T3}(u)\Big\},\non \\ \label{TT}
 \en
where $u$ is the momentum fraction and $\xi \equiv (1-u)-u =1- 2 u$.
Here $\phi_S$, $\phi_{A,T\|}$ and $\phi_{A,T\perp}$ are the leading
twist-$2$ LCDAs, and the others contain contributions from
higher-twist operators. 
The leading twist LCDAs are normalized as
 \be
 \int^1_0 du \phi_{S,A}(u) = 1,\label{normal} \\
 \int^1_0 du \xi \phi_{T}(u) = 1 \label{normalT}
 \en
and can be 
parametrized 
as the so-called $\xi$ moments,
 \be
 \langle \xi^n\rangle=\int^1_{-1} d\xi~\xi^n \phi(\xi).\label{ximoment}
 \en

To disentangle the twist-$2$ LCDAs from higher-twist operators in Eqs.
(\ref{S}) $\sim$ (\ref{TT}), a twist-$2$ contribution of the
relevant nonlocal operator $\bar q (z) \Gamma  q(-z)$ must be
derived. In the case of $\Gamma = \gamma^\mu (\gamma_5)$, the leading
twist-$2$ contribution contains contributions from the operators which
are fully symmetric in Lorentz indices \cite{Ball2,BB}:
 \be
 [\bar q (-z)\gamma^\mu (\gamma_5)
 q(z)]_2=\sum^\infty_{n=0} \frac{1}{n!}\bar q
 (0)\bigg\{\frac{(z\cdot \widehat{D})^n}{n+1} \gamma^\mu +
 \frac{n (z\cdot \widehat{D})^{n-1}}{n+1}\widehat{D}^\mu {\not
 \!z}\bigg\}(\gamma_5)q(0), \label{twist2expand}
 \en
where $\widehat{D}=\overrightarrow{D}-\overleftarrow{D}$ and
$\overrightarrow{D}=\overrightarrow{\partial}-ig B^a (\lambda^a/2)$.
The sum can be represented in terms of a nonlocal operator,
 \be
 [\bar q (-z)\gamma^\mu (\gamma_5)
 q(z)]_2= \int^1_0 dt \frac{\partial}{\partial z_\mu} \bar q (-t z) \not
 \!z (\gamma_5) q(t z). \label{t1}
 \en
Taking the matrix element between the vacuum and the $p$-wave meson
state, we obtain
 \be
 \langle 0 |[\bar q (-z)\gamma^\mu
 q(z)]_2|S(P)\rangle &=& f_S \int^1_0 du
 \phi_S(u) \Bigg\{p^\mu e^{i\xi p z} +(P^\mu -p^\mu)\int^1_0 dte^{i\xi
 t p z}\Bigg\}, \label{Sphi}\\
 \langle 0 |[\bar q (-z)\gamma^\mu \gamma_5
 q(z)]_2|A(P,\epsilon_{\lambda=0})\rangle &=& if_A M_A \int^1_0 du
 \phi_{A\|}(u) \Bigg\{p^\mu \frac{\epsilon z}{p z} e^{i\xi p z}\non
 \\
 &&~~~~~~~~~~~~~~~~~~+\left(\epsilon^\mu-p^\mu \frac{\epsilon z}{p z}\right)\int^1_0 dte^{i\xi
 t p z}\Bigg\},\label{Aphi}\\
 \langle 0 |[\bar q (-z)\gamma^\mu
 q(z)]_2|T(P,\epsilon_{\lambda=0})\rangle &=& f_T M^2_T \int^1_0 du
 \phi_{T\|}(u) \Bigg\{p^\mu \frac{\epsilon^{\bullet\bullet}}{(p z)^2} e^{i\xi p
 z}\non \\
 &&~~~~~~~~~~~~~~~~~~+2\left(\frac{\epsilon^{\mu\bullet}}{p z}-p^\mu
 \frac{\epsilon^{\bullet\bullet}}{(p z)^2}\right)\int^1_0 dte^{i\xi
 t p z}\Bigg\}. \label{Tphi}
 \en
For the derivations in Eqs. (\ref{Sphi}) $\sim$ (\ref{Tphi}), we refer to Ref. \cite{Ball2},
which dealt with the vector meson state. 
We can use Eq. (\ref{twist2expand}), and then expand
the right-hand sides of Eqs. (\ref{Sphi}) $\sim$ (\ref{Tphi}), as follows:
 \be
 &&~~~\sum^\infty_{n=0} \frac{1}{n!}\langle 0|\bar q
 (0)\bigg\{\frac{(z\cdot \widehat{D})^n}{n+1} \gamma^\mu +
 \frac{n (z\cdot \widehat{D})^{n-1}}{n+1}\widehat{D}^\mu {\not
 \!z}\bigg\}q(0)|S(P)\rangle\non \\
 &=& f_S \sum^\infty_{n=0}
 \frac{i^n}{n!}\int^1_0 du \phi_S(u) (\xi p z)^n \Bigg\{p^\mu +(P^\mu -p^\mu)\int^1_0
 dt t^n\Bigg\},\label{Sn}\\
 &&~~~\sum^\infty_{n=0} \frac{1}{n!}\langle 0|\bar q
 (0)\bigg\{\frac{(z\cdot \widehat{D})^n}{n+1} \gamma^\mu +
 \frac{n (z\cdot \widehat{D})^{n-1}}{n+1}\widehat{D}^\mu {\not
 \!z}\bigg\}\gamma_5q(0)|A(P,\epsilon)\rangle\non \\
 &=& if_A M_A \sum^\infty_{n=0}
 \frac{i^n}{n!}\int^1_0 du \phi_{A\|}(u) (\xi p z)^n \Bigg\{p^\mu \frac{\epsilon z}{p z}
  +\left(\epsilon^\mu -p^\mu \frac{\epsilon z}{p z}\right)\int^1_0
 dt t^n\Bigg\},\label{An}\\
 &&~~~\sum^\infty_{n=0} \frac{1}{n!}\langle 0|\bar q
 (0)\bigg\{\frac{(z\cdot \widehat{D})^n}{n+1} \gamma^\mu +
 \frac{n (z\cdot \widehat{D})^{n-1}}{n+1}\widehat{D}^\mu {\not
 \!z}\bigg\}q(0)|T(P,\epsilon)\rangle\non \\
 &=& f_T M^2_T \sum^\infty_{n=0}
 \frac{i^n}{n!}\int^1_0 du \phi_{T\|}(u) (\xi p z)^n \Bigg\{p^\mu \frac{\epsilon^{\bullet\bullet}}{(p z)^2}
  +2\left(\frac{\epsilon^{\mu\bullet}}{p z} -p^\mu \frac{\epsilon^{\bullet\bullet}}{(p z)^2}\right)\int^1_0
 dt t^n\Bigg\}, \label{Tn}
 \en
respectively. Picking $n=0$ in Eqs. (\ref{Sn}) and (\ref{An}), we
obtain
 \be
 \langle 0|\bar q (0) \gamma^\mu q(0)|S(P)\rangle &=& f_S P^\mu
 \int^1_0 du \phi_S (u), \label{S0}\\
 \langle 0|\bar q (0) \gamma^\mu \gamma_5 q(0)|A(P,\epsilon_{\lambda=0})\rangle &=& i f_A M_A \epsilon^\mu
 \int^1_0 du \phi_{A\|} (u).\label{A0}
 \en
Note that the tensor meson cannot be produced by the $V-A$ current.
We then pick $n=1$ in Eq. (\ref{Tn}) and obtain
 \be
 \frac{1}{2}\langle 0|\bar q (0) (\gamma^\mu
 z\cdot\widehat{D}
 +\not\!z \widehat{D}^\mu) q(0)|T(P,\epsilon_{\lambda=0})\rangle &=& f_T M_T^2
 \epsilon^{\mu\bullet} \int^1_0 du \xi\phi_{T\|} (u).\label{T1}
 \en
From the normalization Eq. (\ref{normal}), we have $\langle 0|\bar q
\gamma^\mu \gamma_5 q|^3A_1(P,\epsilon)\rangle = i f_{^3A_1}
 M_{^3A_1} \epsilon^\mu$ which is consistent with the results of Ref. \cite{Yang3}.

Next, we consider the case of $\Gamma = \sigma_{\mu\nu}(\gamma_5)$,
where the leading twist-$2$ contribution contains contributions from
the operators:
 \be
 [\bar q (-z)\sigma^{\mu\nu} (\gamma_5)
 q(z)]_2&=&\sum^\infty_{n=0} \frac{1}{n!}\bar q
 (0)\bigg\{\frac{(z\cdot \widehat{D})^n}{2 n+1} \sigma^{\mu\nu} +
 \frac{n (z\cdot \widehat{D})^{n-1}}{2 n+1}\widehat{D}^\mu
 \sigma^{\bullet\nu}\non \\
 &&\qquad\qquad\quad+\frac{n (z\cdot \widehat{D})^{n-1}}{2
 n+1}\widehat{D}^\nu
 \sigma^{\mu\bullet}\bigg\}(\gamma_5)q(0). \label{twist2expandsigma}
 \en
The sum can also be represented in terms of nonlocal operators:
 \be
 [\bar q (-z)\sigma^{\mu\nu} (\gamma_5) q(z)]_2= \int^1_0 dt \left[\frac{\partial}
 {\partial z_\mu} \bar q (-t^2 z) \sigma^{\bullet\nu} (\gamma_5) q(t^2 z)+z_\alpha \frac{\partial}
 {\partial z_\nu} \bar q (-t^2 z) \sigma^{\mu\alpha} (\gamma_5) q(t^2 z)\right]. \label{t2}
 \en
Taking the matrix element between the vacuum and the axial-vector
and tensor meson state, we obtain:
 \be
 &&\langle 0 |[\bar q (-z)\sigma^{\mu\nu} \gamma_5
 q(z)]_2|A(P,\epsilon_{\lambda=\pm 1})\rangle \non \\
 &=& f^\perp_A \int^1_0 du
 \Bigg\{\phi_{A\perp}(u) \bigg[{\cal S^{\mu\nu}}  e^{i\xi p z}
 +\bigg((\epsilon^\mu P^\nu-\epsilon^\nu P^\mu)-
 {\cal S^{\mu\nu}}\bigg)\int^1_0 dte^{i\xi t^2 p z}\bigg]\non \\
 &&\quad\quad\qquad+\bigg(h_{A\|}(u)-\phi_{A\perp}(u)\bigg)
 \Bigg[{\cal T^{\mu\nu}} e^{i\xi p z}+\bigg({\cal U^{\mu\nu}}-
 {\cal T^{\mu\nu}}\bigg)\int^1_0 dte^{i\xi t^2 p
 z}\Bigg]\Bigg\},\label{Aphisigma}\\
 &&\langle 0 |[\bar q (-z)\sigma^{\mu\nu}
 q(z)]_2|T(P,\epsilon_{\lambda=\pm 1})\rangle \non \\
 &=& if^\perp_T M_T \int^1_0 du
 \Bigg\{\phi_{T\perp}(u) \bigg[{\cal S'^{\mu\nu}}  e^{i\xi p z}
 +\bigg(\frac{2(\epsilon^{\mu\bullet} P^\nu-\epsilon^{\nu\bullet} P^\mu)}{p z}-
 3{\cal S'^{\mu\nu}}\bigg)\int^1_0 dte^{i\xi t^2 p z}\bigg]\non \\
 &&\quad\quad\qquad+\bigg(h_{T\|}(u)-\phi_{T\perp}(u)\bigg)
 \Bigg[{\cal T'^{\mu\nu}} e^{i\xi p z}+\bigg(\frac{2{\cal U'^{\mu\nu}}}{p z}-3
 {\cal T'^{\mu\nu}}\bigg)\int^1_0 dte^{i\xi t^2 p
 z}\Bigg]\Bigg\},\label{Tphisigma}
 \en
where
 \be
 {\cal S^{\mu\nu}}&=&\frac{1}{2}\Bigg[(\epsilon^\mu P^\nu-\epsilon^\nu
 P^\mu)
 -(\epsilon^{\mu}_\perp z^\nu -\epsilon^{\nu}_\perp z^\mu)\frac{M_A^2}{2 p z}\Bigg],\non \\
 {\cal T^{\mu\nu}}&=&\frac{\epsilon z M_A^2}{2 (p z)^2} (p^\mu z^\nu-p^\nu z^\mu)
 ,\qquad\qquad
 {\cal U^{\mu\nu}}=\frac{M_A^2}{p z} (\epsilon^\mu z^\nu-\epsilon^\nu z^\mu)
 ,\non \\
 {\cal S'^{\mu\nu}}&=&\frac{1}{2 p z}\Bigg[(\epsilon^{\mu\bullet} P^\nu-\epsilon^{\nu\bullet} P^\mu)
 -(\epsilon^{\mu\bullet}_\perp z^\nu -\epsilon^{\nu\bullet}_\perp z^\mu)\frac{M_T^2}{2 p z}\Bigg],\non \\
 {\cal T'^{\mu\nu}}&=&\frac{\epsilon^{\bullet\bullet} M_T^2}{2 (p z)^3} (p^\mu z^\nu-p^\nu z^\mu)
 ,\qquad\qquad
 {\cal U'^{\mu\nu}}=\frac{M_T^2}{p z} (\epsilon^\mu z^\nu-\epsilon^\nu
 z^\mu).
 \en
The derivations of Eqs. (\ref{Aphisigma}) and (\ref{Tphisigma}) are
shown in Appendix B of Ref. \cite{JHEPhwang}. In contrast to Eqs. (\ref{Sphi}) $\sim$
(\ref{Tphi}), the twist-$2$ LCDAs do not disentangle entirely from
the higher twists in Eqs. (\ref{Aphisigma}) and (\ref{Tphisigma}).
Taking the product with $\epsilon_{\perp\mu} z_\nu$ and
$\epsilon_{\perp\mu\bullet} z_\nu$ in Eqs. (\ref{Aphisigma}) and
(\ref{Tphisigma}), respectively, we obtain
 \be
 \langle 0 |[\bar q (-z)\sigma^{\mu\bullet}\epsilon_{\perp\mu}
  \gamma_5 q(z)]_2|A(P,\epsilon_{\lambda=\pm 1})\rangle &=& f^\perp_A \int^1_0 du
 \phi_{A\perp}(u) \frac{1}{2}(\epsilon\cdot \epsilon_\perp P z)\bigg[e^{i\xi p z}
 +\int^1_0 dte^{i\xi t^2 p z}\bigg],\non \\ \label{Aphisigmadis}\\
 \langle 0 |[\bar q (-z)\sigma^{\mu\bullet} \epsilon_{\perp\mu \bullet}
 q(z)]_2|T(P,\epsilon_{\lambda=\pm 1})\rangle
 &=& if^\perp_T M_T\int^1_0 du
 \phi_{T\perp}(u) \frac{1}{2} \epsilon^{\mu\bullet} \epsilon_{\perp \mu\bullet} \bigg[ e^{i\xi p z}
 +\int^1_0 dte^{i\xi t^2 p z}\bigg].\non \\ \label{Tphisigmadis}
 \en
Then, we use Eq. (\ref{twist2expandsigma}) and expand the right-hand
sides of Eqs. (\ref{Aphisigmadis}) and (\ref{Tphisigmadis}) as
 \be
 &&~~~\sum^\infty_{n=0} \frac{1}{n!}\langle 0|\bar q
 (0)\frac{(n+1)(z\cdot \widehat{D})^n}{2 n+1}
 \sigma^{\mu\bullet}\epsilon_{\perp\mu} \gamma_5
 q(0)|A(P,\epsilon_{\lambda=\pm 1})\rangle \non \\
 &=& f^\perp_A  \sum^\infty_{n=0}
 \frac{i^n}{n!}\int^1_0 du \phi_{A\perp}(u) \frac{1}{2}(\epsilon\cdot \epsilon_\perp P z)
 (\xi p z)^n \Bigg[1+\int^1_0 dt t^{2n}\Bigg],\label{Ansigma}
 \en
 \be
 &&~~~\sum^\infty_{n=0} \frac{1}{n!}\langle 0|\bar q
 (0)\frac{(n+1)(z\cdot \widehat{D})^n}{2 n+1}
 \sigma^{\mu\bullet}\epsilon_{\perp\mu}  q(0)|T(P,\epsilon_{\lambda=\pm 1})\rangle\non \\
 &=& if^\perp_T M_T \sum^\infty_{n=0}
 \frac{i^n}{n!}\int^1_0 du \phi_{T\perp}(u) \frac{1}{2} \epsilon^{\mu\bullet} \epsilon_{\perp\mu \bullet}
 (\xi p z)^n \Bigg[1+\int^1_0 dt t^{2n}\Bigg], \label{Tnsigma}
 \en
Picking $n=0$ and $n=1$ in Eqs. (\ref{Ansigma}) and (\ref{Tnsigma}), respectively, we
obtain
 \be
 \langle 0|\bar q(0)\sigma^{\mu\bullet}\epsilon_{\perp\mu} \gamma_5
 q(0)|A(P,\epsilon_{\lambda=\pm 1})\rangle &=& f^\perp_A\int^1_0 du \phi_{A\perp}(u) (\epsilon\cdot
 \epsilon_\perp P z),\label{Ansigmafm}\\
 \langle 0|\bar q (0)(z\cdot \widehat{D})\sigma^{\mu\bullet}\epsilon_{\perp\mu\bullet}  q(0)|T(P,\epsilon_{\lambda=\pm 1})\rangle
 &=& f^\perp_T M_T
 \int^1_0 du \xi \phi_{T\perp}(u)(\epsilon^{\mu\bullet} \epsilon_{\perp\mu \bullet}P z). \label{Tnsigmafm}
 \en
\subsection{Heavy Quark Framework}
In general, the theoretical description of meson properties relies
on the bound state models with a relativistic normalization:
 \be
 \langle M(P',\epsilon')|M(P,\epsilon)\rangle=2P^0 (2\pi)^3
 \delta^3(P'-P)\delta_{\epsilon\epsilon'}.\label{normnorm}
 \en
At low energies, however, these models have little connection to the
fundamental theory of QCD. The reliable predictions are often
made based on symmetries. A well-known example is HQS
\cite{Neubert}, which arises since the Compton wavelength, $1/m_Q$,
of a heavy quark bound inside a hadron is much smaller than a
typical hadronic distance (about $1$ fm), and $m_Q$ is unimportant
for the low-energy properties of the state. For a heavy-light meson
system, it is more natural to use velocity $v^\mu$ instead of
momentum variables. Then, it is appropriate to work with a
mass-independent normalization of a heavy-light meson state:
 \be
 \langle \widehat{M}(v',\hat{\epsilon}')|\widehat{M}(v,\hat{\epsilon})\rangle=2v^0 (2\pi)^3
 \delta^3(\bar{\Lambda}
 v'-\bar{\Lambda}v)\delta_{\hat{\epsilon}\hat{\epsilon}'}, \label{normHQ}
 \en
where $\bar \Lambda=M-m_Q$ is the so-called residual center mass of
a heavy-light meson. The relation between these two bound states is
 \be
 |M(P,\epsilon)\rangle =\sqrt{M}|\widehat{M}(v,\hat{\epsilon})\rangle. \label{states}
 \en
In addition, the heavy quark field can be expanded as \cite{Neubert}
 \be
 Q(x)=e^{-im_Q v\cdot x}\left[1+\frac{1}{iv\cdot D+2 m_Q-i\varepsilon}
 i\not\!\! D_\perp\right]h_v(x),
 \en
where $h^*_v(x)$ is a field describing a heavy antiquark with
velocity $v$. Then, the current $\bar q \Gamma Q$ can be represented
as
 \be
 \bar q \Gamma Q =\bar q \Gamma \left(1+\frac{i\not\!\! D_\perp}{2 m_Q}
 +\cdot\cdot\cdot\right)h_v. \label{currentE}
 \en
Substituting Eqs. (\ref{states}) and (\ref{currentE}) into the
definitions of LCDAs, Eqs. (\ref{S}) $\sim$ (\ref{TT}) yield
 \be
 \langle 0|\bar q (z) \gamma^\mu h_v (-z)|\widehat{S}(v)\rangle &=& F_S \int^\infty_0 d\omega~e^{i\omega v z}\left[v^\mu
  \Phi_S(\omega)+z^\mu \frac{1}{2v z}G_S(\omega)\right], \label{Sh}\\
 \langle 0|\bar q (z) \gamma^\mu \gamma_5 h_v (-z)|\widehat{A}(v,\hat{\epsilon}_{\lambda=0})\rangle
 &=&i F_A  \int^\infty_0 d\omega~e^{i\omega v z}\Big[v^\mu \frac{\hat{\epsilon} z}{v z}
 \Phi_{A\|}
 (\omega) +\hat{\epsilon}^\mu_{\perp}G_{A\perp}(\omega)\non \\
 &&~~~~~~~~~~~~~~~~~~~~~~- z^\mu \frac{\hat{\epsilon} z}{2 (v z)^2}
 G_{A3}(u)\Big], \label{ALh}
 \en
 \be
 &&\langle 0|\bar q (z) \sigma^{\mu\nu} \gamma_5 h_v (-z)|\widehat{A}(v,\hat{\epsilon}_{\lambda=\pm 1})\rangle
 =F^\perp_A \int^\infty_0 d\omega~e^{i\omega v z}\Big[(\hat{\epsilon}^\mu_{\perp} v^\nu-\hat{\epsilon}^\nu_{\perp} v^\mu)
 \Phi_{A\perp} (u)\non \\
 &&\qquad\qquad\qquad\qquad\qquad\qquad+(v^\mu z^\nu-v^\nu z^\mu)\frac{ \hat{\epsilon} z}{(v z)^2}H_{A\|}(\omega)
 +(\hat{\epsilon}^\mu_{\perp} z^\nu-\hat{\epsilon}^\nu_{\perp} z^\mu) \frac{1}{2 v z}
 H_{A3}(\omega)\Big],\non \\ \label{ATh}\\
 &&\langle 0|\bar q (z) \gamma^\mu h_v (-z)|\widehat{T}(v,\hat{\epsilon}_{\lambda=0})\rangle
 =F_T \int^\infty_0 d\omega~e^{i\omega v z}\Big[v^\mu \frac{\hat{\epsilon}^{\bullet\bullet}}{(v z)^2}
 \Phi_{T\|}(\omega) +\frac{\hat{\epsilon}^{\mu\bullet}_\perp}{v z}G_{T\perp}(\omega)\non \\
 &&\qquad\qquad\qquad\qquad\qquad\qquad- z^\mu \frac{\hat{\epsilon}^{\bullet\bullet}}
 {2 (v z)^3} G_{T3}(\omega)\Big],\label{TLh} \\
 &&\langle 0|\bar q (z) \sigma^{\mu\nu} h_v (-z)|\widehat{T}(v,\hat{\epsilon}_{\lambda=\pm 1})\rangle
 =iF^\perp_T \int^\infty_0 d\omega~e^{i\omega v
 z}\Big[\frac{(\hat{\epsilon}^{\mu\bullet}_\perp
 v^\nu-\hat{\epsilon}^{\bullet\nu}_\perp
 v^\mu)}{v z}
 \Phi_{T\perp} (\omega)\non \\
 &&\qquad\qquad\qquad\qquad\qquad+(v^\mu z^\nu-v^\nu z^\mu)\frac{
 \hat{\epsilon}^{\bullet\bullet}}{(v z)^3}H_{T\|}(\omega)
 +(\hat{\epsilon}^{\mu\bullet}_\perp z^\nu-\hat{\epsilon}^{\bullet\nu}_\perp z^\mu) \frac{1}{2 (v z)^2}
 H_{T3}(\omega)\Big],\non \\ \label{TTh}
 \en
where $F_M=\sqrt{M}f_M$, $\Phi_i (\omega)=\phi_i(u)/M$, and $\omega$
was first introduced in Ref. \cite{CZL} as the product of the
longitudinal momentum fraction $u$ of the light (anti)quark and the
mass of heavy meson $M$, namely, $\omega=u M$. Following a similar
process, the leading twist LCDAs are obtained as
\be
 \langle 0|\bar q (0) \gamma^\mu h_v(0)|\widehat{S}(v)\rangle &=& F_S v^\mu
 \int^\infty_0 d\omega \Phi_S (\omega), \label{S0h}\\
 \langle 0|\bar q (0) \gamma^\mu \gamma_5 h_v(0)|\widehat{A}(v,\hat{\epsilon}_{\lambda=0})\rangle &=& i F_A  \hat{\epsilon}^\mu
 \int^\infty_0 d\omega \Phi_{A\|} (\omega),\label{A0h}\\
 \frac{1}{2}\langle 0|\bar q (0) (\gamma^\mu
 z\cdot\widehat{D}
 +\not\!z \widehat{D}^\mu) h_v(0)|\widehat{T}(v,\hat{\epsilon}_{\lambda=0})\rangle &=& F_T \hat{\epsilon}^{\mu\bullet} \int^\infty_0 d\omega \Phi_{T\|} (\omega),\label{T1h} \\
 \langle 0|\bar q(0)\sigma^{\mu\bullet}\hat{\epsilon}_{\perp\mu} \gamma_5
 h_v(0)|\widehat{A}(v,\hat{\epsilon}_{\lambda=\pm 1})\rangle &=& F^\perp_A(\hat{\epsilon}\cdot
 \hat{\epsilon}_\perp v z)\int^\infty_0 d\omega \Phi_{A\perp}(\omega),\label{Ansigmafmh}\\
 \langle 0|\bar q (0)(z\cdot \widehat{D})\sigma^{\mu\bullet}\hat{\epsilon}_{\perp\mu\bullet}  h_v(0)|\widehat{T}(v,\hat{\epsilon}_{\lambda=\pm 1})\rangle
 &=& F^\perp_T \hat{\epsilon}^{\mu\bullet} \hat{\epsilon}_{\perp\mu \bullet}
 \int^\infty_0 d\omega \Phi_{T\perp}(\omega). \label{Tnsigmafmh}
 \en
\section{General Formulism in LFQM }
\subsection{General Framework}

A meson bound state, consisting of a quark $q_1$ and an antiquark
$\bar q_2$ with total momentum $P$ and spin $J$, can be written as
(see, for example, Ref. \cite{CCH1})
 \be
 |M(P, L, J)\rangle =\int &&\{d^3k_1\}\{d^3k_2\} ~2(2\pi)^3
 \delta^3(\tilde P -\tilde k_1-\tilde k_2)~\non\\
 &&\times \sum_{\lambda_1,\lambda_2}
 \Psi^{JJ_z}_{LS}(\tilde k_1,\tilde k_2,\lambda_1,\lambda_2)~
 |q_1(k_1,\lambda_1) \bar q_2(k_2,\lambda_2)\rangle,\label{lfmbs}
 \en
where $k_1$ and $k_2$ are the on-mass-shell light-front momenta,
 \be
 \tilde k=(k^+, k_\bot)~, \quad k_\bot = (k^1, k^2)~,
 \quad k^- = \frac{m_q^2+k_\bot^2}{k^+},
 \en
and
 \be
 &&\{d^3k\} \equiv \frac{dk^+d^2k_\bot}{2(2\pi)^3}, \nonumber \\
 &&|q(k_1,\lambda_1)\bar q(k_2,\lambda_2)\rangle
 = b^\dagger_{\lambda_1}(k_1)d^\dagger_{\lambda_2}(k_2)|0\rangle,\\
 &&\{b_{\lambda'}(k'),b_{\lambda}^\dagger(k)\} =
 \{d_{\lambda'}(k'),d_{\lambda}^\dagger(k)\} =
 2(2\pi)^3~\delta^3(\tilde k'-\tilde k)~\delta_{\lambda'\lambda}.
 \nonumber
 \en
In terms of the light-front relative momentum variables $(u,
\kappa_\bot)$ are defined by
 \be
 && k^+_1=(1-u) P^{+}, \quad k^+_2=u P^{+}, \nonumber \\
 && k_{1\bot}=(1-u) P_\bot+\kappa_\bot, \quad k_{2\bot}=u
 P_\bot-\kappa_\bot.
 \en

The momentum-space wave function $\Psi^{JJ_z}_{LS}$ for a
$^{2S+1}L_J$ meson can be expressed as
 \be
 \Psi^{ JJ_z}_{LS}(\tilde k_1,\tilde k_2,\lambda_1,\lambda_2)
 = \frac{1}{\sqrt N_c}\la L S; L_z S_z|L S;J J_z\ra
 R^{SS_z}_{\lambda_1\lambda_2}(u,\kappa_\bot)~ \varphi_{LL_z}(u,
 \kappa_\bot),\label{Psi}
 \en
where $\varphi_{LL_z}(u,\kappa_\bot)$ describes the momentum
distribution of the constituent quarks in the bound state with the
orbital angular momentum $L$, $\langle L S; L_z S_z|L S;J
J_z\rangle$ as the corresponding Clebsch-Gordan coefficient and
$R^{SS_z}_{\lambda_1\lambda_2}$ constructing a state of definite spin
($S,S_z$) out of light-front helicity ($\lambda_1,\lambda_2$)
eigenstates. Explicitly,
 \be
 R^{SS_z}_{\lambda_1 \lambda_2}(u,\kappa_\bot)
 =\sum_{s_1,s_2} \langle \lambda_1|
  {\cal R}_M^\dagger(1-u,\kappa_\bot, m_1)|s_1\rangle
 \langle \lambda_2|{\cal R}_M^\dagger(u,-\kappa_\bot, m_2)
 |s_2\rangle \langle \frac{1}{2}\,\frac{1}{2};s_1
 s_2|\frac{1}{2}\frac{1}{2};SS_z\rangle,
 \en
where $|s_i\rangle$ are the usual Pauli spinors, and ${\cal R}_M$ is
the Melosh transformation operator~\cite{Jaus1}:
 \be
 \langle s|{\cal R}_M (u_i,\kappa_\bot,m_i)|\lambda\rangle
 &=&\frac{m_i+u_i M_0
 +i\vec \sigma_{s\lambda}\cdot\vec \kappa_\bot \times
 \vec n}{\sqrt{(m_i+u_i M_0)^2 + \kappa^{2}_\bot}},
 \en
with $u_1=1-u$, $u_2=u$, and 
$\vec n =(0,0,1)$ is a unit vector in the $\hat {z}$-direction. In
addition,
 \be
 M_0^2&=&(e_1+e_2)^2=\frac{m_1^2+\kappa^2_\bot}{u_1}+\frac{m_2^2+\kappa^2_\bot}{u_2},
 \non \\
 e_i&=&\sqrt{m^2_i+\kappa^2_\perp+\kappa^2_z},\quad \frac{e_1-\kappa_z}{e_1+e_2}=1-u, \quad\frac{e_2+\kappa_z}{e_1+e_2}=u,\non
 \en
where $\kappa_z$ is the relative momentum in the $\hat{z}$ direction and
can be written as
 \be \label{eq:Mpz}
  \kappa_z=\frac{u M_0}{2}-\frac{m^2_2+\kappa^2_\perp}{2 u M_0}.
 \en
$M_0$ is the invariant mass of $q\bar q$ and is generally
different from the mass $M$ of a meson which satisfies $M^2=P^2$. This
is due to the fact that the meson, quark and antiquark cannot be
simultaneously on-shell. We normalize the meson state as
 \be
 \langle M(P',J',J'_z)|M(P,J,J_z)\rangle = 2(2\pi)^3 P^+
 \delta^3(\tilde P'- \tilde P)\delta_{J'J}\delta_{J'_z J_z}~,
 \label{wavenor}
 \en
in order that
 \be
 \int \frac{du\,d^2\kappa_\bot}{2(2\pi)^3}~\varphi^{\prime*}_{L^\prime
 L^\prime_z}(u,\kappa_\bot)
 \varphi_{LL_z}(u,\kappa_\bot)
 =\delta_{L^\prime,L}~\delta_{L^\prime_z,L_z}.
 \label{momnor}
 \en
Explicitly, we have
 \be
 \varphi_{1L_z}=\kappa_{L_z} \varphi_p,
 \en
where $\kappa_{L_z=\pm1}=\mp(\kappa_{\bot x}\pm i \kappa_{\bot
y})/\sqrt2$, $\kappa_{L_z=0}=\kappa_{z}$ are proportional to the
spherical harmonics $Y_{1L_z}$ in momentum space, and $\varphi_p$ is the
distribution amplitude of the $p$-wave meson. In general, for any
function $F(|\vec{\kappa}|)$, $\varphi_p(u,\kappa_\perp)$ has the form of
 \be
 \varphi_p(u,\kappa_\perp)=N \sqrt{\frac{d\kappa_z}{du}}F(|\vec{\kappa}|),\label{F}
 \en
where
 \be
 \frac{d\kappa_z}{du}=\frac{e_1 e_2}{u (1-u) M_0}\label{Jac}
 \en
is the Jacobian of transformation from $(u,\kappa_\perp)$ to $\vec{\kappa}$ and the normalization factor $N$ is determined from Eq.~(\ref{momnor}).


In the case of a $p$-wave meson state, it is more convenient to use
the covariant form of $R^{SS_z}_{\lambda_1\lambda_2}$
\cite{Jaus1,CCH2,cheung}:
 \be
 \langle 1 S; L_z S_z|1 S;J J_z\rangle\, k_{L_z} \,R^{SS_z}_{\lambda_1\lambda_2}(u,\kappa_\bot)
 &=&\frac{\sqrt{k_1^+ k_2^+}}{\sqrt2~{\widetilde M_0}(M_0+m_1+m_2)}\non \\
 &&\times\bar u(k_1,\lambda_1)(\not\!\!\bar P+M_0)\Gamma_{^{2S+1}\!P_J}
 v(k_2,\lambda_2), \label{covariantp}
 \en
where
 \be
 &&\widetilde M_0\equiv\sqrt{M_0^2-(m_1-m_2)^2},\qquad\quad \bar
 P\equiv k_1+k_2,\non \\
 &&\bar u(k,\lambda) u(k,\lambda')=\frac{2
 m}{k^+}\delta_{\lambda,\lambda'},\qquad\quad \sum_\lambda u(k,\lambda)
 \bar u(k,\lambda)=\frac{\not\!k +m}{k^+},\non \\
 &&\bar v(k,\lambda) v(k,\lambda')=-\frac{2
 m}{k^+}\delta_{\lambda,\lambda'},\qquad\quad \sum_\lambda v(k,\lambda)
 \bar v(k,\lambda)=\frac{\not\!k -m}{k^+}.
 \en
For the scalar, axial-vector, and tensor mesons, we have
 \be
 \Gamma_{^3\!P_0}&=&\frac{1}{\sqrt3}\left(\not\!\!K-\frac{K\cdot
 \bar P}{M_0}\right),\non\\
 \Gamma_{^1\!P_1}&=&\epsilon\cdot K \gamma_5,\non\\
 \Gamma_{^3\!P_1}&=&\frac{1}{\sqrt2}\left((\not\!\!K-\frac{K\cdot
 \bar P}{M_0})\not\!\epsilon-\epsilon\cdot K\right)\gamma_5, \non\\
 \Gamma_{^3\!P_2}&=& \epsilon_{\mu\nu}\gamma^\mu(- K^\nu),\label{Gamma}
 \en
where $K\equiv (k_2-k_1)/2$ and
 \be
 &&\epsilon^\mu_{\lambda=\pm 1} =
 \left[\frac{2}{ P^+} \vec \epsilon_\bot (\pm 1) \cdot
 \vec P_\bot,\,0,\,\vec \epsilon_\bot (\pm 1)\right],\non \\
 &&\vec \epsilon_\bot (\pm 1)=\mp(1,\pm i)/\sqrt{2}, \non\\
 &&\epsilon^\mu_{\lambda=0}=\frac{1}{M_0}\left(\frac{-M_0^2+P_\bot^2}{
 P^+},P^+,P_\bot\right).   \label{polcom}
 \en
Note that the polarization tensor of a tensor meson satisfies the
relations $\epsilon_{\mu\nu}=\epsilon_{\nu\mu}$ and
$\epsilon_{\mu\nu}\bar P^\mu = \epsilon^\mu_\mu = 0$. Equations
(\ref{covariantp}) and (\ref{Gamma}) can be further reduced by the
applications of equations of motion on spinors:
 \be
 \langle 1 S; L_z S_z|1 S;J J_z\rangle\, k_{L_z}
 \,R^{SS_z}_{\lambda_1\lambda_2}(u,\kappa_\bot)
 =\frac{\sqrt{k_1^+ k_2^+}}{\sqrt2~{\widetilde M_0}}
 \bar u(k_1,\lambda_1)\Gamma'_{^{2S+1}\!P_J}
 v(k_2,\lambda_2), \label{covariantfurther}
 \en
where
 \be
 \Gamma'_{^3\!P_0}&=&-\frac{\widetilde M_0^2}{2\sqrt{3} M_0},\non\\
 \Gamma'_{^1\!P_1}&=&\epsilon\cdot K \gamma_5,\non\\
 \Gamma'_{^3\!P_1}&=&\frac{-1}{2\sqrt{2} M_0}\left(\not\!\epsilon
 \widetilde M_0^2-2\epsilon\cdot K(m_1-m_2)\right)\gamma_5, \non\\
 \Gamma'_{^3\!P_2}&=& \epsilon_{\mu\nu}\left(\gamma^\mu+\frac{2 K^\mu}
 {M_0+m_1+m_2}\right)(- K^\nu).\label{Gammap}
 \en

Next, the matrix elements of Eqs. (\ref{S0}), (\ref{A0}),
(\ref{T1}), (\ref{Ansigmafm}), and (\ref{Tnsigmafm}) will be
calculated within the LFQM, and the relevant leading twist LCDAs are
extracted. For the scalar meson state, we substitute Eqs.
(\ref{lfmbs}), (\ref{Psi}), and (\ref{covariantfurther}) into Eq.
(\ref{S0}) to obtain
 \be
 \langle 0|\bar q_2 \gamma^\mu q_1|S(P)\rangle&=&N_c\int\{d^3
 k_1\}\sum_{\lambda_1,\lambda_2}\Psi_{LS}^{JJ_z}(k_1,k_2,\lambda_1,\lambda_2)
 \langle 0 |\bar q_2 \gamma^\mu q_1|q_1\bar q_2\rangle\non \\
 &=& -\sqrt{N_c}\int\{d^3 k_1\}\frac{\sqrt{k_1^+ k_2^+}}{\sqrt2~{\widetilde
 M_0}}\varphi_p{\rm Tr}\bigg[\gamma^\mu \Bigg(\frac{\not\!k_1+m_1}{k_1^+}\Bigg)\frac{\widetilde M_0^2}{2\sqrt{3}
 M_0}\Bigg(\frac{-\not\!k_2+m_2}{k^+_2}\Bigg)\bigg]\non \\
 &=&f_S P^\mu \int du \phi(u).
 \en
For the ``good" component, $\mu=+$, the leading twist LCDA $\phi_S$
can be extracted as
 \be
 \phi_S(u) = \frac{\sqrt{2 N_c}}{f_S}\int \frac{d^2
 \kappa_\perp}{2(2\pi)^3}\frac{[u m_1-(1-u) m_2]\widetilde
 M_0}{\sqrt{3u(1-u)}M_0}\varphi_p(u,\kappa_\perp).\label{Su}
 \en
A similar process can be used for the axial-vector and tensor mesons
which correspond to Eqs. (\ref{A0}), (\ref{Ansigmafm}), and
(\ref{T1}), (\ref{Tnsigmafm}), respectively, and the leading twist
LCDAs are extracted as
 \be
 \phi_{^3\!A_1 \|}(u)&=&-\frac{\sqrt{3}}{f_{^3\!A_1}}\int \frac{d^2
 \kappa_\perp}{2(2\pi)^3}\frac{\varphi_p(u,\kappa_\perp)}{\sqrt{u(1-u)} M_0 \widetilde M_0}\Bigg[
 \frac{\widetilde M_0^2}{2 M_0}(\widetilde M_0^2-4 m_1 m_2)+\widetilde M_0^2 \kappa_z (1-2 u)\non \\
 &&\qquad-2 \kappa_z (m_1-m_2) (u m_1+(1-u) m_2)\Bigg],\label{3A1}\\
 \phi_{^1\!A_1\|}(u) &=& \frac{2\sqrt{6}}{f_{^1\!A_1}}\int \frac{d^2
 \kappa_\perp}{2(2\pi)^3}\frac{u m_1+(1-u) m_2}{\sqrt{u(1-u)}\widetilde M_0}\varphi_p(u,\kappa_\perp)\kappa_z,\label{1A1}\\
 \phi_{^3\!A_1 \perp}(u)&=&-\frac{\sqrt{3}}{f^\perp_{^3\!A_1}}\int \frac{d^2
 \kappa_\perp}{2(2\pi)^3}\frac{\varphi_p(u,\kappa_\perp)}{\sqrt{u(1-u)}M_0}\Bigg[\widetilde M_0(u m_1-(1-u)m_2) +\frac{\kappa_\perp^2}{\widetilde M_0}(m_1-m_2)\Bigg], \label{3A1p}\\
 \phi_{^1\!A_1 \perp}(u)&=&\frac{\sqrt{6}}{f^\perp_{^1\!A_1}}\int \frac{d^2
 \kappa_\perp}{2(2\pi)^3}\frac{\kappa^2_\perp}{\sqrt{u(1-u)}\widetilde M_0}\varphi_p(u,\kappa_\perp),\label{1A1p} \\
 \phi_{T\|}(u) &=& \frac{\sqrt{6}}{f_{T}}\int \frac{d^2
 \kappa_\perp}{2(2\pi)^3}\frac{ \varphi_p(u,\kappa_\perp) }{\sqrt{u(1-u)}\widetilde M_0}\Bigg[\kappa_z\frac{\widetilde M_0^2}{M_0}\non \\
 &&\qquad+(2 \kappa^2_z-\kappa^2_\perp)\Big((1-2 u)+2 \frac{(u m_1-(1-u) m_2)}{M_0+m_1+m_2}\Big)\Bigg],
 \label{Tkperp}\\
 \phi_{T\perp}(u) &=& \frac{2\sqrt{6}}{f^\perp_{T}}\int \frac{d^2
 \kappa_\perp}{2(2\pi)^3}\frac{\varphi_p(u,\kappa_\perp)}{\sqrt{u(1-u)}\widetilde M_0}\Bigg[\kappa_z (u m_1+(1-u) m_2)-\frac{\kappa^2_\perp}{2 M_0}(m_1-m_2)\non \\
 &&\qquad +\frac{2\kappa^2_\perp}
 {M_0+m_1+m_2}\kappa_z\Bigg].\label{Tk}
 \en
\subsection{Heavy Quark Framework}
If one takes $m_1 =m_Q \to \infty$, that is, the heavy quark limit
in the heavy-light meson, then two inequalities, $m_Q \simeq M_0 \gg
m_2$, $\kappa_\perp$ and $u \to 0$, are obtained. 
The exact form of $\Phi_M$ can be derived by the
redefinition of the meson bound state. Let us consider the bound
states of heavy mesons in the heavy quark limit:
 \begin{eqnarray}  \label{hqslfb}
  |\widehat{M}(v;L,J)\rangle &=& \int \{d^3q\}\{d^3k_2\} 2(2\pi)^3 \delta^3(
    \overline{\Lambda}\tilde{v}-\tilde{q}-\tilde{k}_2) \nonumber \\
    &\times& \sum_{\lambda_1,\lambda_2}
    \widehat{\Psi}^{JJ_z}_{LS} (\omega,\kappa_{\bot},\lambda_1,\lambda_2)
    b_v^\dagger(q, \lambda_1) d^\dagger (k_2, \lambda_2)|0\rangle,
 \end{eqnarray}
where $q=k_1-m_Q v$ is the residual momentum of the heavy quark. Operators $b_v^\dagger(q,\lambda_1)$ create a heavy quark with
\begin{eqnarray}
    \{ b_v (q,\lambda_1), ~ b_{v'}^\dagger (q',\lambda'_1) \}
        &=&2 (2\pi)^3 \delta_{vv'}\delta^3(\tilde{q}-\tilde{q}')
        \delta_{\lambda_1 \lambda'_1}.
\end{eqnarray}
The variable $\omega$ and the relative transverse and longitudinal momenta, $\kappa_\bot$ and
$\kappa_z$, are obtained by
 \begin{equation}
 \omega=e_2+\kappa_z,~~\kappa_\bot=k_{2\bot}-\omega v_\bot,~~\kappa_z
 ={\omega\over{2}}-{m_2^2+\kappa_\bot^2\over{2 \omega}}.\label{kzH}
 \end{equation}
The momentum-space wave function $\widehat{\Psi}^{JJ_z}_{LS}$ can be
expressed as
 \be
 \widehat{\Psi}^{JJ_z}_{LS}(\omega,\kappa_{\bot},\lambda_1,\lambda_2)
 = \frac{1}{\sqrt N_c}
 \langle LS;L_z S_z|LS,J J_z\rangle\widehat{R}^{SS_z}_{\lambda_1\lambda_2}(\omega,\kappa_\bot)~
 \widehat{\varphi}_{LL_z}(\omega, \kappa_\bot),\label{hatPsi}
 \en
where
 \begin{equation}  \label{spin}
  \langle LS;L_z S_z|LS,J J_z\rangle\kappa_{L_z}\widehat{R}^{SS_z}(\omega, \kappa_{\bot}, \lambda_1, \lambda_2)
    =\frac{k_2^+}{\sqrt{2}
        \sqrt{(\omega+m_2)^2+\kappa^2_\perp}} ~ \bar u (v,\lambda_1)
        \hat{\Gamma} v(k_2,\lambda_2)
 \end{equation}
with 
 \be
 \hat{\Gamma}_{^3P_0}&=&-\frac{1}{\sqrt{3}}(v\cdot k_2+m_2),\non \\
 \hat{\Gamma}_{^1P_1}&=& \hat{\epsilon}\cdot k_2\gamma_5,\non \\
 \hat{\Gamma}_{^3P_1}&=&-\frac{1}{\sqrt{2}}\Big[(v\cdot k_2+m_2)\not\!\hat{\epsilon}-\hat{\epsilon}\cdot k_2\Big]\gamma_5,\non \\
 \hat{\Gamma}_{^3P_2}&=&-\hat{\epsilon}_{\mu\nu}\gamma^\mu k_2^\nu,
 \en
and
 \be
 &&\hat{\epsilon}^\mu_{\lambda=\pm 1} =
 \left[\frac{2}{ v^+} \vec \epsilon_\bot (\pm 1) \cdot
 \vec v_\bot,\,0,\,\vec \epsilon_\bot (\pm 1)\right],\non\\
 &&\hat{\epsilon}^\mu_{\lambda=0}=\left(\frac{-1+v_\bot^2}{
 v^+},v^+,v_\bot\right).  \label{Hpolcom}
 \en
$u(v,\lambda_1)$ is the spinor for the heavy quark,
 \begin{equation}
    \sum_\lambda u(v,\lambda)\overline{u}(v,\lambda)
        = \frac{{\not \! v}+1}{v^+}.
\end{equation}
The normalization of the heavy meson bound states can then be given
by
 \begin{equation}  \label{nmc2}
    \langle \widehat{M}(v',J',J'_z) |\widehat{M}(v,J,J_z)\rangle = 2(2\pi)^3 v^+
        \delta^3(\overline{\Lambda}v'-\overline{\Lambda}v)
        \delta_{JJ'} \delta_{J_zJ'_z},
 \end{equation}
which not only leads to 
Eq. (\ref{states}), but also to the space part
$\widehat{\varphi}_{LL_z}(\omega,\kappa_\bot^2)$ (called the
light-front wave function) in Eq. (\ref{hqslfb}) which has the
following wave-function normalization condition:
 \begin{equation} \label{nwf}
   \int \frac{d\omega d^2\kappa_\bot} {2(2\pi)^3}
        |\widehat{\varphi}_{LL_z} (\omega, \kappa^2_\bot)|^2 = 1,
 \end{equation}
where $\widehat{\varphi}_{LL_z}=\kappa_{L_z}\widehat{\varphi}_p$.
In principle, the heavy quark dynamics are completely described by
HQET, which is given by the $1/m_Q$ expansion of the heavy quark QCD
Lagrangian:
\begin{equation}
    {\cal L} = \overline{Q} (i \not \! \! D - m_Q) Q \nonumber
    = \sum_{n=0}^\infty \Bigg({1 \over 2m_Q}
        \Bigg)^n {\cal L}_n.
        \label{hqcdl}
\end{equation}
Therefore, $|\widehat{M}(v;L,J)\rangle$ and
$\widehat{\varphi}_p (\omega,\kappa^2_{\bot})$ are
determined by the leading Lagrangian ${\cal L}_0=\bar h_v iv\cdot D
h_v$. 
From the normalization conditions of Eqs. (\ref{momnor}) and
(\ref{nwf}), we obtain the relation between wave functions
$\varphi_p(u,\kappa^2_\perp)$ and
$\widehat{\varphi}_p(\omega,\kappa^2_\perp)$:
 \be \label{scale}
 \varphi_p(u,\kappa^2_\perp)=\sqrt{M}\widehat{\varphi}_p(\omega,\kappa^2_\perp).
 \en
In addition, in the heavy quark limit $(m_1 \rightarrow\infty)$, the
heavy quark spin $s_Q$ decouples from the other degrees of
freedom so that $s_Q$ and the total angular momentum of the
light antiquark $j$ are separately good quantum numbers.
Hence, it is more convenient to use the $L_J^j=P_2^{3/2}$, $P_1^{3/2}$,
$P_1^{1/2}$ and $P_0^{1/2}$ basis. It is obvious that the first and the last of these
states are $^3P_2$ and $^3P_0$ , respectively, while \cite{IW}
 \be
 |P^{1/2}_1\rangle&=&\frac{1}{\sqrt{3}}|^1P_1\rangle-\sqrt{\frac{2}{3}}|^3P_1\rangle,\non \\
 |P^{3/2}_1\rangle&=&\sqrt{\frac{2}{3}}|^1P_1\rangle+\frac{1}{\sqrt{3}}|^3P_1\rangle.
 \en
In terms of $P_1^{1/2}$ and $P_1^{3/2}$ states, the relevant vertex functions
read
 \be
 \hat{\Gamma}_{P_1^{1/2}}&=&\frac{1}{\sqrt{3}}(v\cdot k_2+m_2)\not\!\hat{\epsilon}\gamma_5,\non \\
 \hat{\Gamma}_{P_1^{3/2}}&=&-\frac{1}{\sqrt{6}}[(v\cdot k_2+m_2)\not\!\hat{\epsilon}-3 \hat{\epsilon}\cdot k_2]\gamma_5.
 \en
Next, the matrix elements of Eqs.
(\ref{S0h}),(\ref{A0h}),(\ref{T1h}),(\ref{Ansigmafmh}) and
(\ref{Tnsigmafmh}) can be calculated, and the relevant leading twist
LCDAs are extracted as
 \be
 F_S\Phi_S(\omega)&=&F_{A^{1/2}_1}\Phi_{A^{1/2}_1\|}(\omega)=F^\perp_{A^{1/2}_1}
 \Phi_{A^{1/2}_1\perp}(\omega)\non \\
 &=&\sqrt{2}\int \frac{d^2\kappa_\perp}{2 (2\pi)^3}\frac{\omega-m_2}{\omega}\sqrt{(\omega+m_2)^2+\kappa_\perp^2}\hat{\varphi}_p
 (\omega,\kappa_\perp^2),\label{A12S}\\
 F_{A^{3/2}_1}\Phi_{A^{3/2}_1\|}(\omega)&=& \int \frac{d^2\kappa_\perp}{2 (2\pi)^3}\frac{\hat{\varphi}_p
 (\omega,\kappa_\perp^2)}{\sqrt{(\omega+m_2)^2+\kappa_\perp^2}}[\omega^2-m^2_2-3 \kappa^2_\perp+2 \kappa^2_z+2 m_2 \kappa_z],\label{A32}\\
 F^\perp_{A^{3/2}_1}\Phi_{A^{3/2}_1\perp}(\omega)&=& \int \frac{d^2\kappa_\perp}{2 (2\pi)^3}\frac{\hat{\varphi}_p
 (\omega,\kappa_\perp^2)}{\sqrt{(\omega+m_2)^2+\kappa_\perp^2}}[2 \kappa^2_z- \kappa_\perp^2],\label{A32p}\\
 F_{T}\Phi_{T\|}(\omega)&=&\sqrt{6}\int \frac{d^2\kappa_\perp}{2 (2\pi)^3}\frac{\hat{\varphi}_p
 (\omega,\kappa_\perp^2)}{\sqrt{(\omega+m_2)^2+\kappa_\perp^2}}\Bigg[\kappa_z\frac{(\omega+m_2)^2+\kappa_\perp^2}
 {\omega}+2 \kappa^2_z-\kappa^2_\perp\Bigg],\label{T32}\\
 F^\perp_{T}\Phi_{T\perp}(\omega)&=& \sqrt{6}\int \frac{d^2\kappa_\perp}{2 (2\pi)^3}\frac{\hat{\varphi}_p
 (\omega,\kappa_\perp^2)}{\sqrt{(\omega+m_2)^2+\kappa_\perp^2}}[\omega^2-m^2_2-2 \kappa^2_\perp+ m_2 \kappa_z],\label{T32p}
 \en
where we have denoted the $P_1^{1/2}$ and $P_1^{3/2}$ states by $A^{1/2}_1$ and
$A^{3/2}_1$, respectively. In the literature, we read that \cite{YOPR,VD} HQS requires
 \be
 f_{A^{1/2}_1}=f_S, \quad f_{A^{3/2}_1}=0.\label{HQS}
 \en
These relations can
be understood from the fact that $(P_0^{1/2},P_1^{1/2})$ and
$(P_1^{3/2},P_2^{3/2})$ form two doublets in the heavy quark limit and that the
tensor meson cannot be induced from the $V-A$ current. From Eq. (\ref{A12S}), it is easy to find that
 \be
 F_S=F_{A^{1/2}_1}=F^\perp_{A^{1/2}_1}, \quad \Phi_S(\omega)=\Phi_{A^{1/2}_1\|}(\omega)=\Phi_{A^{1/2}_1\perp}(\omega). \label{SArelation}
 \en
Combining the former of Eq. (\ref{SArelation}) and the scaling $F_M=\sqrt{M} f_M$, we obtain: $f_S=f_{A^{1/2}_1}=f^\perp_{A^{1/2}_1}$, which are consistent with the former of Eq. (\ref{HQS}). On the
other hand, if wave function $\hat{\varphi}_p$
has a similar form to Eq. (\ref{F}),
 \be
 \hat{\varphi}_p(\omega,\kappa_\perp)=N'\sqrt{\frac{d\kappa_z}{d\omega}}F(|\vec{\kappa}|),\label{Fform}
 \en
where $N'$ is the normalization constant and $F(|\vec{\kappa}|)$ is a
function of $|\vec{\kappa}|$, then, Eqs. (\ref{A32}) and (\ref{A32p}) can be shown as
$F_{A^{3/2}_1}=F^\perp_{A^{3/2}_1}=0$ or $f_{A^{3/2}_1}=f^\perp_{A^{3/2}_1}=0$, which are also consistent with the latter of Eq. (\ref{HQS}). The derivations are shown in Appendix
A. In addition, the tensor meson can be created through the $V-A$
currents with covariant derivatives [see Eqs. (\ref{T1}) and
(\ref{Tnsigmafm})]. Thus, we can study its decay constant $F_T$ and
$F^\perp_T$ here. From Appendix A, we find that Eqs. (\ref{T32}) and
(\ref{T32p}) lead to $F_T=F^\perp_T=0$ and then $f_T=f^\perp_T=f_{A^{3/2}_1}=f^\perp_{A^{3/2}_1}=0$, which is consistent with the fact that $(P_1^{3/2},P_2^{3/2})$ forms a doublet in the heavy quark limit.
\section{Numerical results and discussions}
In this section, the decay constants and LCDAs for the $p$-wave states of $D$, $D_s$,
$B$ and $B_s$ systems are studied with
the wave function $\varphi_p(u,\kappa_\perp)$. In principle,
$\varphi_p$ is obtained by solving the light-front QCD bound-state
equation $H_{LF} |M\rangle = M|M\rangle$, which is the familiar
Schr$\ddot{\textrm{o}}$dinger equation in ordinary quantum mechanics,
and $H_{LF}$ is the light-front Hamiltonian. However, except in some
simple cases, achieving the full solution has remained a challenge. There are
several popular phenomenological light-front
momentum distribution amplitudes which have been employed to describe
various hadronic structures in the literature. A widely used one is
the Gaussian type. If 
 \be\label{FF}
 F^g(|\vec \kappa|)&=&{\rm exp}\bigg(-\frac{|\vec \kappa|^2}{2
 \beta^2}\bigg),
 \en
then the corresponding wave functions are 
 \be
 \varphi^g_p(u,\kappa_\perp)&=&4\sqrt{\frac{2}{\beta^2}}\bigg(\frac{\pi}
 {\beta^2}\bigg)^{3/4}\sqrt{\frac{e_1 e_2}{u (1-u) M_0}}~{\rm
 exp}\bigg[-\frac{\kappa_\perp^2+(\frac{u M_0}{2}-\frac{m^2_2+\kappa^2_\perp}{2 u M_0})^2}{2
 \beta^2}\bigg].\label{Gaussian1s}
 \en
Prior to numerical
calculations, the parameters $m_1$, $m_2$, and $\beta$, which
appeared in the wave function, have to first be determined. Here, we
use the parameters obtained in the ISGW2 model \cite{ISGW2}, an update
of the ISGW quark model \cite{ISGW} for semileptonic meson decays.
They are listed in Table I.
\begin{table}[ht!]
\caption{\label{tab:parameter} The input parameters $m_1$, $m_2$, and $\beta$ (in units of GeV) in the
Gaussian-type wave function Eq. (\ref{Gaussian1s}).}
\begin{ruledtabular}
\begin{tabular}{cccc|cccc}
 $m_{u,d}$ & $m_s$ & $m_c$ & $m_b$ & $\beta_{cu}$ &  $\beta_{cs}$ & $\beta_{bu}$ & $\beta_{bs}$ \\\hline
$0.33$ & $0.55$   & $1.82$ & $5.20$ & $0.33$&  $0.38$ & $0.35$ & $0.41$
\end{tabular}
\end{ruledtabular}
\end{table}
Next, we use the parameters in Table I to evaluate the relevant decay constants, Eqs. (\ref{Su})$\sim$ (\ref{Tk}), for the $p$-wave heavy mesons. Note that the decay constant of the scalar and axial vector mesons are calculated by normalization of Eq. (\ref{normal}), and that of the tensor mesons are calculated by normalization of Eq. (\ref{normalT}).
The results, which compare with other theoretical evaluations, are listed in Table II.
\begin{table}[ht!]
 \caption{\label{tab:decay constant} Mesonic decay constants (in units of MeV) obtained in this work and other theoretical evaluations.}
 \begin{ruledtabular}
 \begin{tabular}{ccccccc}
 $^{2 S+1} L_J (L^j_J)$ & $^3P_0$ & $^3P_1$ & $^1P_1$ & $^3P_2$  & $P_1^{1/2}$  & $P^{3/2}_1$  \\ \hline
 $f_{cu}$ & $78$   & $-113$ & $45$ & $-50$ & $118$   & $-29$  \\
 $f_{cu}$ \cite{CCH2} & $86$ & $-127$ & $45$ & & $130$ & $-36$  \\
 $f_{cu}$ \cite{VD} & $139\pm30$ &  &  & & $251\pm37$ & $77\pm18$  \\
 $f_{cu}$ \cite{Wang} & $133$ &  $-211$ & $72$  & &  &   \\
 $f^\perp_{cu}$ &    & $-127$ & $80$ & $-48$ & $150$   & $-8$ \\ \hline
 $f_{cs}$ & $66$ & $-123$ & $38$ & $-65$ & $123$   & $-40$  \\
 $f_{cs}$ \cite{CCH2} & $71$ & $-121$ & $38$ & & $122$ & $-38$ \\
 $f_{cs}$ \cite{VD} & $110\pm18$ &  &  & & $233\pm31$ & $87\pm19$ \\
 $f_{cs}$ \cite{Wang} & $112$ &  $-219$ & $62$ & &  &   \\
 $f_{cs}^\perp$ &  & $-107$ & $87$ & $-62$ & $138$   & $9$ \\ \hline
 $f_{bu}$ & $73$   & $-73$ & $42$ & $-17$ & $84$   & $-8$ \\
 $f_{bu}$ \cite{CCH2} & $112$ & $-123$ & $68$  & & $140$ & $-15$ \\
 $f_{bu}$ \cite{VD} & $162\pm24$ &  &   & & $206\pm29$ & $32\pm10$ \\
 $f_{bu}$ \cite{Wang} & $145$ & $-150$ & $76$ & &  &   \\
 $f_{bu}^\perp$ &  & $-119$ & $52$ & $-17$ & $127$   & $-26$ \\ \hline
 $f_{bs}$ & $75$   & $-82$ & $43$ & $-25$ & $92$   & $-12$ \\
 $f_{bs}$ \cite{VD} & $146\pm19$ & & & & $196\pm26$ & $36\pm10$\\
 $f_{bs}$ \cite{Wang} & $140$ & $-157$ & $76$ & &  &   \\
 $f_{bs}^\perp$ &  & $-122$ & $58$ & $-25$ & $133$   & $-23$ \\
 \end{tabular}
 \end{ruledtabular}
 \end{table}
We find two things: the first is $f_{A^{1/2}_1}^{(\perp)}\gg f_{A^{3/2}_1}^{(\perp)}$, and the second is $f_T$ and $f^\perp_T$ for the charm sector are larger than those for the bottom sector. Both are qualitatively consistent with the results in Section 3B, which were derived from the HQS. In addition, in this work, the ratio of the decay constants $f_{A^{1/2}_1}$ and $f_S$ in the $D_s$ system is
 \be
 f_{A^{1/2}_1}/f_S=1.9.
 \en
This is very close to the predictions of Refs. \cite{VD} and \cite{HK}: $f_{A^{1/2}_1}/f_S=2.12$ and $2.26\pm0.41$, which are based on the mock-meson approach and the factorization hypothesis, respectively.

For these heavy mesons, it is convenient to study the leading twist LCDAs using the $L^j_J$ basis. Thus, the curves of $\phi(\xi)$ for the $p$-wave states in $D$, $D_s$, $B$, and $B_s$ systems are evaluated and shown in Figs. 1 $\sim$ 7.
 \begin{figure}
 \includegraphics*[width=4in]{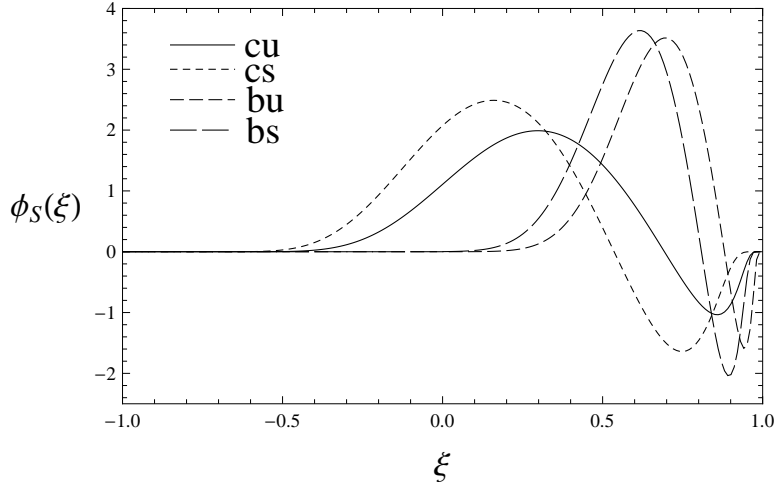}
 \caption{Leading twist-2 LCDAs $\phi_S(\xi)$ of the heavy meson. The
 solid and dotted lines, and short and long dashes correspond to the $D$,
 $D_s$, $B$, and $B_s$ systems, respectively.}
  \label{fig:phis}
 \end{figure}
 \begin{figure}
 \includegraphics*[width=4in]{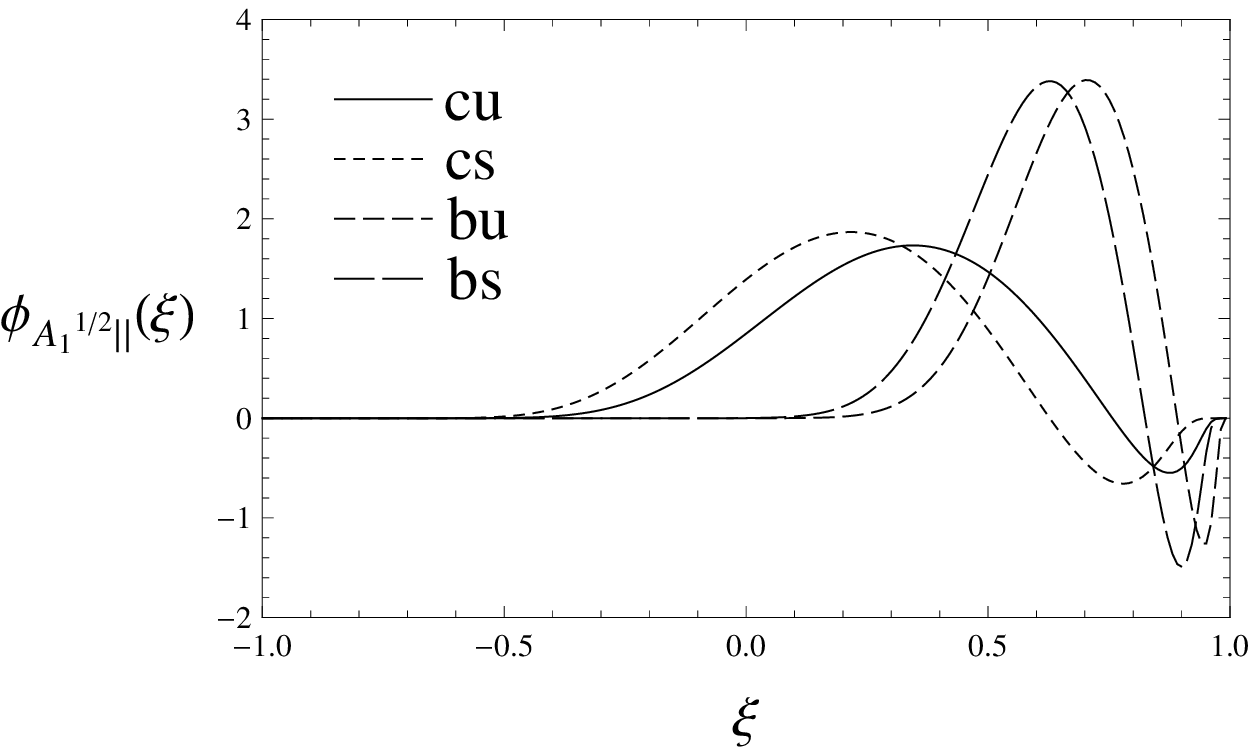}
 \caption{Leading twist-2 LCDAs $\phi_{A^{1/2}_1\|}(\xi)$ of the heavy meson. The
 solid and dotted lines, and short and long dashes correspond to the $D$,
 $D_s$, $B$, and $B_s$ systems, respectively.}
  \label{fig:phiA121}
 \end{figure}
 \begin{figure}
 \includegraphics*[width=4in]{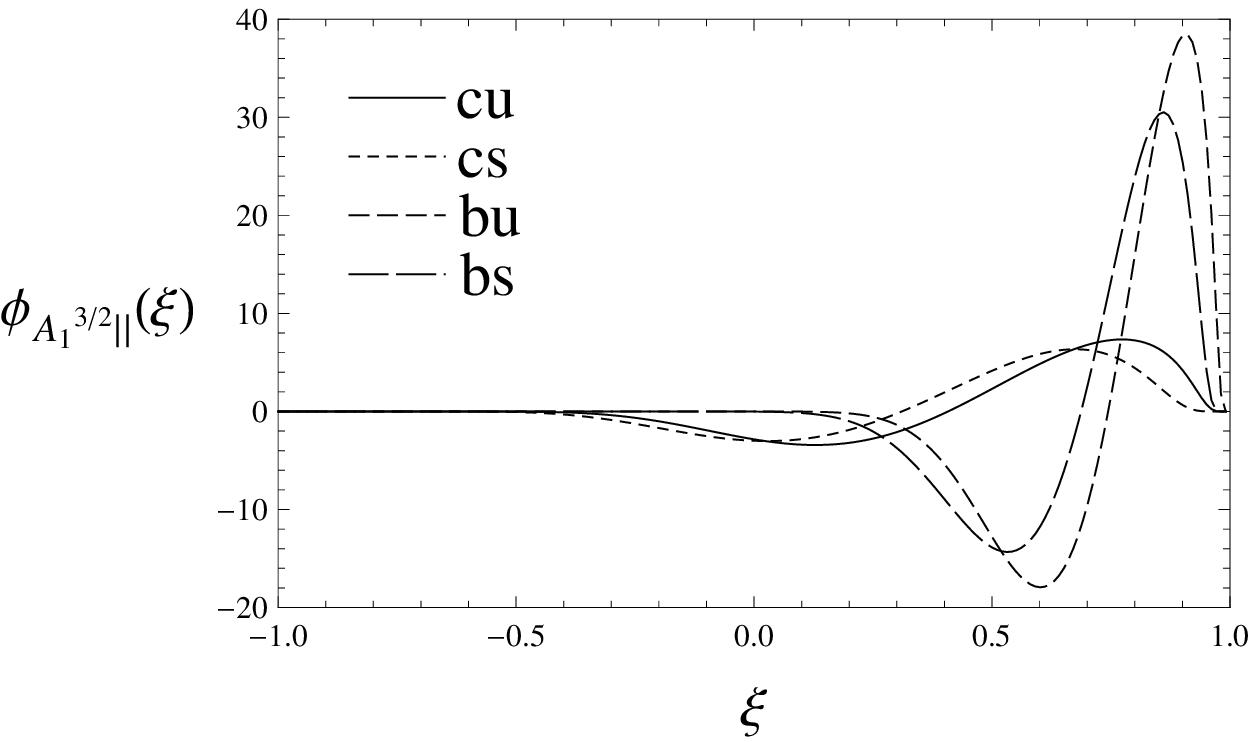}
 \caption{Leading twist-2 LCDAs $\phi_{A^{3/2}_1\|}(\xi)$ of the heavy meson. The
 solid and dotted lines, and short and long dashes correspond to the $D$,
 $D_s$, $B$, and $B_s$ systems, respectively.}
  \label{fig:phiA321}
 \end{figure}
 \begin{figure}
 \includegraphics*[width=4in]{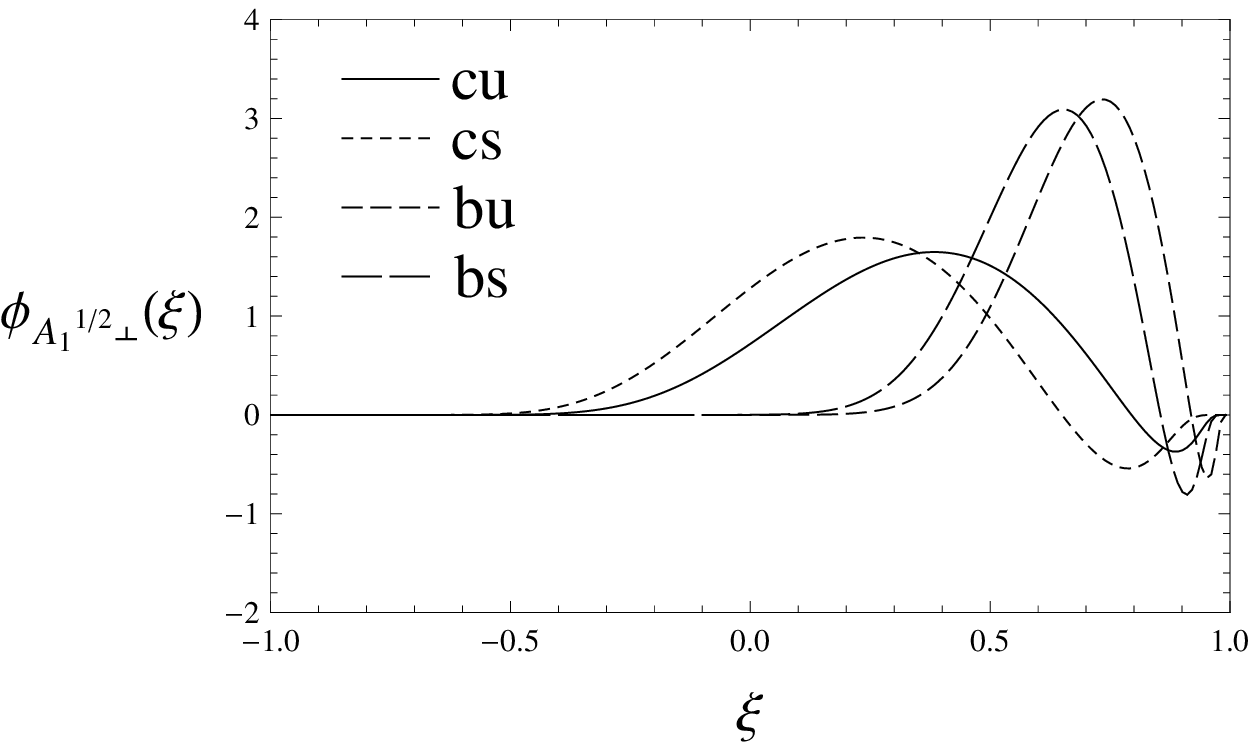}
 \caption{Leading twist-2 LCDAs $\phi_{A^{1/2}_1\perp}(\xi)$ of the heavy meson. The
 solid and dotted lines, and short and long dashes correspond to the $D$,
 $D_s$, $B$, and $B_s$ systems, respectively.}
  \label{fig:phiA121o}
 \end{figure}
 \begin{figure}
 \includegraphics*[width=4in]{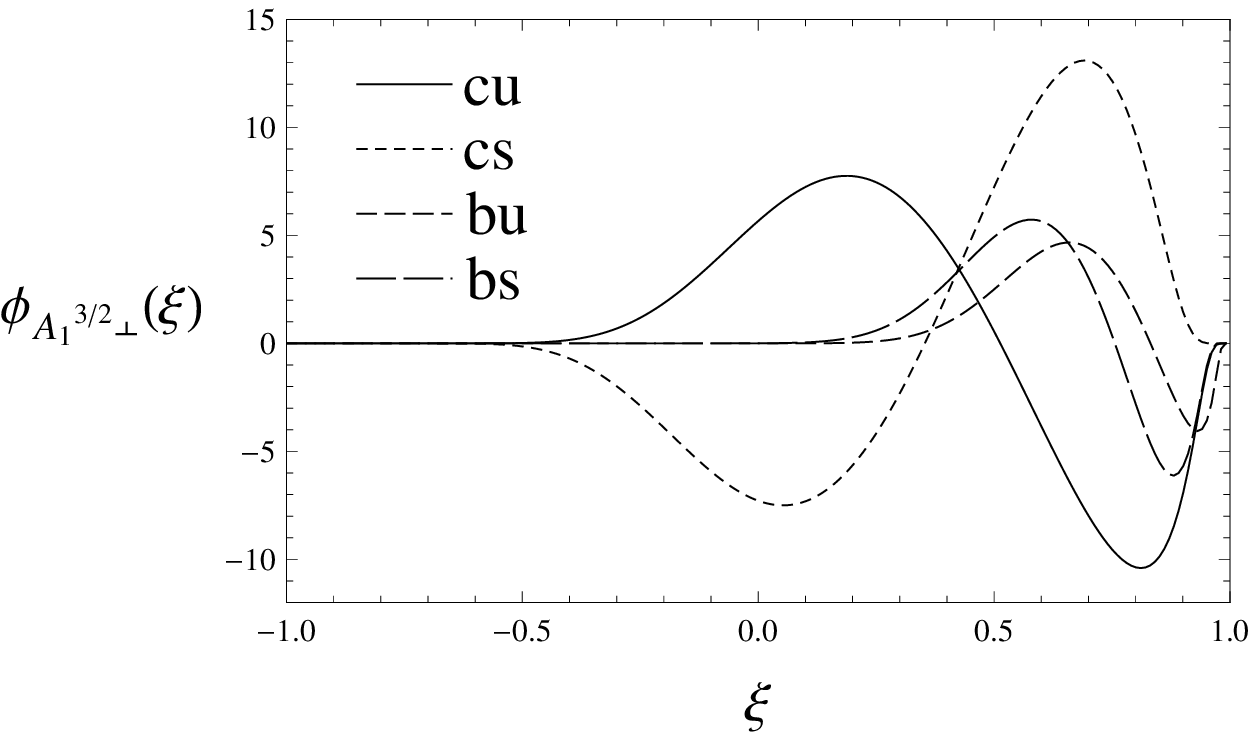}
 \caption{Leading twist-2 LCDAs $\phi_{A^{3/2}_1\perp}(\xi)$ of the heavy meson. The
 solid and dotted lines, and short and long dashes correspond to the $D$,
 $D_s$, $B$, and $B_s$ systems, respectively.}
  \label{fig:phi1A1o}
 \end{figure}
 \begin{figure}
 \includegraphics*[width=4in]{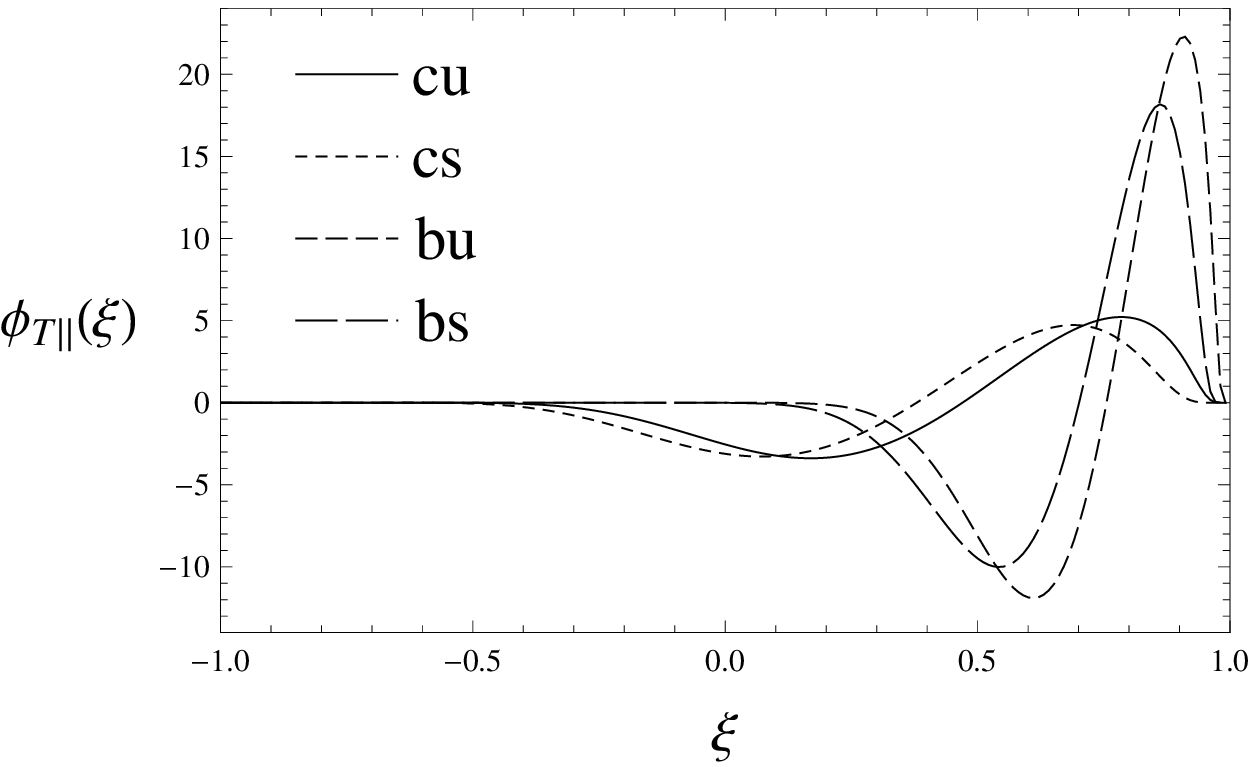}
 \caption{Leading twist-2 LCDAs $\phi_{T\|}(\xi)$ of the heavy meson. The
 solid and dotted lines, and short and long dashes correspond to the $D$,
 $D_s$, $B$, and $B_s$ systems, respectively.}
  \label{fig:phiTp}
 \end{figure}
 \begin{figure}
 \includegraphics*[width=4in]{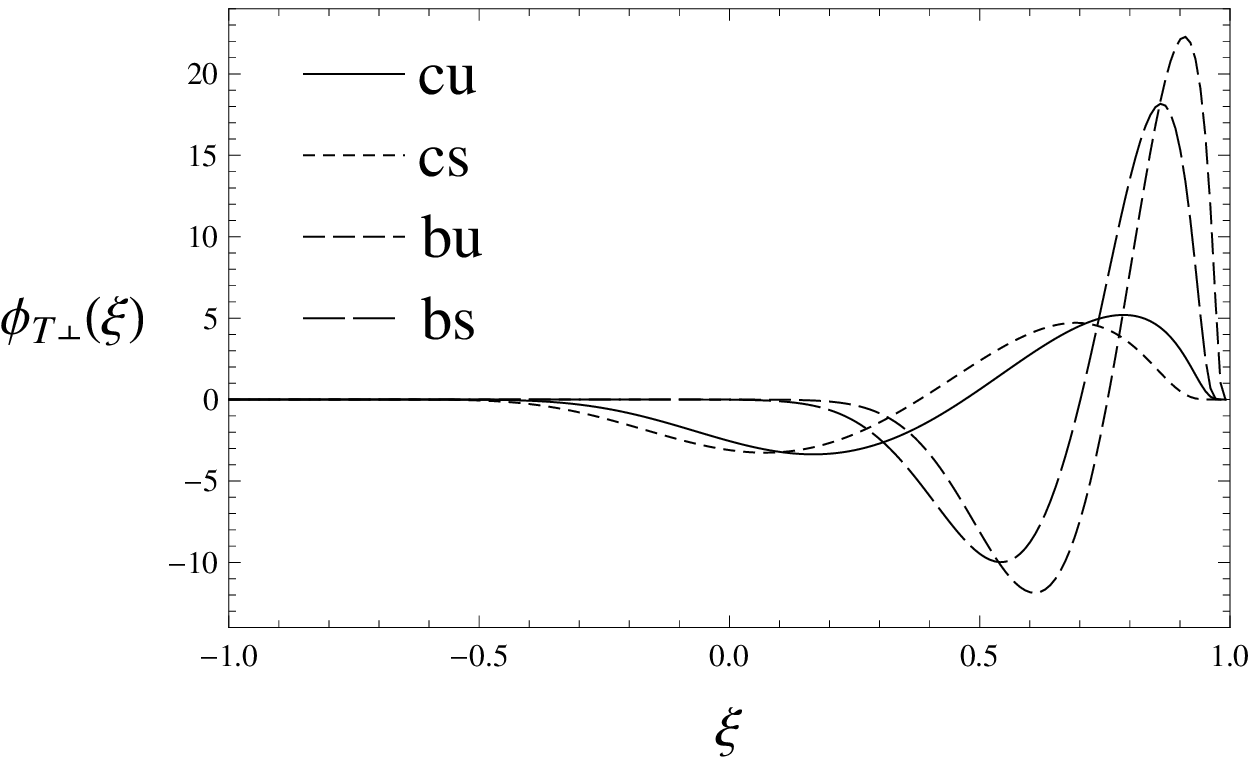}
 \caption{Leading twist-2 LCDAs $\phi_{T\perp}(\xi)$ of the heavy meson. The
 solid and dotted lines, and short and long dashes correspond to the $D$,
 $D_s$, $B$, and $B_s$ systems, respectively.}
  \label{fig:phiTo}
 \end{figure}
From these figures, we find the curves of $\phi_{A^{1/2}_1\|}(\xi)$ and $\phi_{A^{1/2}_1\perp}(\xi)$ are very similar to those of $\phi_S(\xi)$, but are quite different from those of $\phi_{A^{3/2}_1\|}(\xi)$ and $\phi_{A^{3/2}_1\perp}(\xi)$. The scales of Figs. 3, 5, 6, and 7 are much larger than those of the others because the values of $f^{(\perp)}_{A^{3/2}_1}$ and $f_T^{(\perp)}$ are relatively small.
Finally, we parametrize the LCDAs in terms of the first six $\xi$ moments with Eq. (\ref{ximoment}).
The results 
are shown in Tables III, IV, V, and VI.
\begin{table}[ht!]
\caption{\label{tab:ximomentsD} The first four $\xi$ moments of $\phi_M(\xi)$ for the $p$-wave
states of $D$ system.}
\begin{ruledtabular}
\begin{tabular}{cccccccc}
 $M$ & $S$ & $P^{1/2}_1\|$  & $P^{1/2}_1\perp$ & $P^{3/2}_1\|$ & $P^{3/2}_1\perp$ & $T\|$ & $T\perp$ \\ \hline
 $\langle \xi\rangle$   & $0.14$    & $0.25$ & $0.30$  & $1.7$   & $-1.7$ & $1$ & $1$\\
 $\langle \xi^2\rangle$ & $0.0029$  & $0.091$ & $0.13$  & $1.3$  & $-1.5$ & $0.83$ & $0.83$ \\
 $\langle \xi^3\rangle$ & $-0.050$  & $0.023$ & $0.055$  & $1.0$ & $-1.3$ & $0.68$ & $0.68$\\
 $\langle \xi^4\rangle$ & $-0.061$  & $-0.0031$ & $0.023$  & $0.79$ & $-1.1$ & $0.54$ & $0.54$\\
 $\langle \xi^5\rangle$ & $-0.062$  & $-0.015$ & $0.0061$  & $0.62$ & $-0.87$ & $0.43$ & $0.44$\\
 $\langle \xi^6\rangle$ & $-0.057$  & $-0.019$ & $-0.0021$  & $0.50$ & $-0.72$ & $0.35$ & $0.35$\\
\end{tabular}
\end{ruledtabular}
\end{table}
\begin{table}[ht!]
\caption{\label{tab:ximomentsDs} The first four $\xi$ moments of $\phi_M(\xi)$ for the $p$-wave
states of $D_s$ system.}
\begin{ruledtabular}
\begin{tabular}{cccccccc}
 $M$ & $S$ & $P^{1/2}_1\|$  & $P^{1/2}_1\perp$ & $P^{3/2}_1\|$ & $P^{3/2}_1\perp$ & $T\|$ & $T\perp$ \\ \hline
 $\langle \xi\rangle$   & $-0.13$    & $0.093$ & $0.13$  & $1.5$   & $2.9$ & $1$ & $1$\\
 $\langle \xi^2\rangle$ & $-0.14$  & $0.011$ & $0.033$  & $0.96$  & $1.9$ & $0.66$ & $0.66$ \\
 $\langle \xi^3\rangle$ & $-0.14$  & $-0.029$ & $-0.013$  & $0.69$ & $1.4$ & $0.50$ & $0.50$\\
 $\langle \xi^4\rangle$ & $-0.11$  & $-0.032$ & $-0.020$  & $0.48$ & $1.0$ & $0.35$ & $0.36$\\
 $\langle \xi^5\rangle$ & $-0.092$  & $-0.030$ & $-0.022$  & $0.35$ & $0.74$ & $0.26$ & $0.26$\\
 $\langle \xi^6\rangle$ & $-0.072$  & $-0.026$ & $-0.020$  & $0.26$ & $0.55$ & $0.20$ & $0.20$\\
\end{tabular}
\end{ruledtabular}
\end{table}
\begin{table}[ht!]
\caption{\label{tab:ximomentsB} The first four $\xi$ moments of $\phi_M(\xi)$ for the $p$-wave
states of $B$ system.}
\begin{ruledtabular}
\begin{tabular}{cccccccc}
 $M$ & $S$ & $P^{1/2}_1\|$  & $P^{1/2}_1\perp$ & $P^{3/2}_1\|$ & $P^{3/2}_1\perp$ & $T\|$ & $T\perp$ \\ \hline
 $\langle \xi\rangle$   & $0.62$  & $0.63$ & $0.67$   & $2.4$   & $0.50$ & $1$ & $1$\\
 $\langle \xi^2\rangle$ & $0.39$  & $0.41$ & $0.47$  & $2.9$  & $0.22$ & $1.4$ & $1.4$ \\
 $\langle \xi^3\rangle$ & $0.25$  & $0.27$ & $0.33$  & $3.0$ & $0.059$ & $1.6$ & $1.6$\\
 $\langle \xi^4\rangle$ & $0.16$  & $0.18$ & $0.24$  & $2.9$ & $-0.034$ & $1.6$ & $1.6$\\
 $\langle \xi^5\rangle$ & $0.095$  & $0.12$ & $0.18$  & $2.8$ & $-0.088$ & $1.5$ & $1.5$\\
 $\langle \xi^6\rangle$ & $0.054$  & $0.077$ & $0.13$  & $2.6$ & $-0.12$ & $1.4$ & $1.4$\\
\end{tabular}
\end{ruledtabular}
\end{table}
\begin{table}[ht!]
\caption{\label{tab:ximomentsBs} The first six $\xi$ moments of $\phi_M(\xi)$ for the $p$-wave
states of $B_s$ system.}
\begin{ruledtabular}
\begin{tabular}{cccccccc}
 $M$ & $S$ & $P^{1/2}_1\|$  & $P^{1/2}_1\perp$ & $P^{3/2}_1\|$ & $P^{3/2}_1\perp$ & $T\|$ & $T\perp$ \\ \hline
 $\langle \xi\rangle$   & $0.50$  & $0.54$ & $0.58$  & $2.3$   & $0.27$ & $1$ & $1$\\
 $\langle \xi^2\rangle$ & $0.25$  & $0.29$ & $0.35$  & $2.6$  & $-0.057$ & $1.3$ & $1.3$ \\
 $\langle \xi^3\rangle$ & $0.11$  & $0.15$ & $0.22$  & $2.5$ & $-0.20$ & $1.3$ & $1.3$\\
 $\langle \xi^4\rangle$ & $0.035$  & $0.077$ & $0.14$  & $2.3$ & $-0.26$ & $1.3$ & $1.3$\\
 $\langle \xi^5\rangle$ & $-0.0066$  & $0.032$ & $0.085$  & $2.0$ & $-0.28$ & $1.1$ & $1.1$\\
 $\langle \xi^6\rangle$ & $-0.029$  & $0.0046$ & $0.053$  & $1.8$ & $-0.27$ & $1.0$ & $1.0$\\
\end{tabular}
\end{ruledtabular}
\end{table}
We find that the similarities between the longitudinal and transverse projections of the tensor meson are displayed not only in ratio $f_T/f^\perp_T \simeq 1$, but also in the approximations $\langle \xi^i\rangle_{T\|}\simeq \langle \xi^i\rangle_{T\perp}$ for all heavy meson systems.
\section{Conclusions}
This study discussed the leading twist LCDAs of $p$-wave heavy
mesons within the light-front approach. 
These LCDAs have
been displayed in terms of light-front variables $(u,\omega,\kappa_\perp)$
and the relevant decay constants in both general and heavy
quark frameworks. In the heavy quark framework, we analytically found that the
decay constants and LCDAs had the following relations: $f_S=f_{A^{1/2}_1}=f^\perp_{A^{1/2}_1}$,
$\Phi_S(\omega)=\Phi_{A^{1/2}_1\|}(\omega)=\Phi_{A^{1/2}_1\perp}(\omega)$ and $f_T=f^\perp_T=f_{A^{3/2}_1}=f^\perp_{A^{3/2}_1}=0$, which are
consistent with the requirements of HQS-$(P^{1/2}_0,P^{1/2}_1)$ and $(P^{3/2}_1,P^{3/2}_2)$
form two doublets. It was worth noting that we could study $f^{(\perp)}_T$ because the tensor meson was created through the $V-A$ currents with covariant derivatives. In the general framework, we quoted the parameters $m_i$ and $\beta$, which appear in the Gaussian-type wave functions, from the ISGW2 model, and numerically found that: i) an inequality $f_{A^{1/2}_1}^{(\perp)}\gg f_{A^{3/2}_1}^{(\perp)}$ is existent for all systems; ii) the decay constants $f_T$ and $f^\perp_T$ for the charm sector are larger than those for the bottom sector; iii) the ratio $f_{A^{1/2}_1}/f_S=1.9$ for the $D_s$ system is close to the predictions in Ref. \cite{VD} and \cite{HK}; iv) the curves of $\phi_{A^{1/2}_1\|,\perp}(\xi)$ were very similar to those of $\phi_S(\xi)$, but were quite different from those of $\phi_{A^{3/2}_1\|,\perp}(\xi)$ and $\phi_{T\|,\perp}(\xi)$; and v) the ratio $f_T/f^\perp_T\simeq 1$ and approximations $\langle \xi^i\rangle_{T\|}\simeq \langle \xi^i\rangle_{T\perp}$ were satisfied for all heavy meson systems. It is easily realized that the results of i, ii, and iv qualitatively supported the requirements of HQS. Due to the lack of relevant experimental data, the consistencies between our estimations and the predictions of HQS are important.



{\bf Acknowledgements}\\
This work is supported in part by the National Science
Council of R.O.C. under Grant No. NSC-99-2112-M-017-002-MY3.

\appendix
\section{Some useful identities}
We consider an integration as
 \be
 \int\frac{d\omega d^2\kappa_\perp}{2(2\pi)^3}\frac{\hat{\varphi}_p(\omega,\kappa^2_\perp)}{\sqrt{(\omega+m_2)^2+\kappa^2_\perp}} g(m_2,\kappa_z,\kappa^2_\perp).
 \label{A1}
 \en
Substituting Eq. (\ref{Fform}) with Eq. (\ref{A1}), we obtain
 \be
 N'\int\frac{d\omega d^2\kappa_\perp}{2(2\pi)^3}\sqrt{\frac{d\kappa_z}{d\omega}}\frac{F(|\vec{\kappa}|)}
 {\sqrt{(\omega+m_2)^2+\kappa^2_\perp}} g(m_2,\kappa_z,\kappa^2_\perp).
 \label{A2}
 \en
Taking the heavy quark limit for Eq. (\ref{Jac}), we obtain $\sqrt{\frac{d\kappa_z}{d\omega}}=\sqrt{\frac{e_2}{\omega}}$, and Eq. (\ref{A2}) can be rewritten as
 \be
 N'\int\frac{d^3\vec{\kappa}}{2(2\pi)^3}\sqrt{\frac{\omega}{e_2}}\frac{F(|\vec{\kappa}|)}
 {\sqrt{(\omega+m_2)^2+\kappa^2_\perp}} g(m_2,\kappa_z,\kappa^2_\perp).
 \label{A3}
 \en
From Eq. (\ref{kzH}), the variables $\omega$, $e_2$, and $\kappa_z$ have the following relations:
 \be
 \omega=e_2+\kappa_z,\quad \frac{m^2_2+\kappa^2_\perp}{\omega}=e_2-\kappa_z.
 \en
Thus, Eq. (\ref{A3}) can be rewritten as
 \be
 N'\int\frac{d^3\vec{\kappa}}{2(2\pi)^3}\frac{F(|\vec{\kappa}|)}
 {\sqrt{2e_2(e_2+m_2)}} g(m_2,\kappa_z,\kappa^2_\perp).
 \label{A5}
 \en
Besides the function $g$, the only variable in Eq. (\ref{A5}) is $|\hat{\kappa}|$ because $e_2=\sqrt{m^2_2+|\vec{\kappa}|^2}$. Therefore, if we designate $g$ to some specific function in Eq. (\ref{A1}), the integration can be made aware by symmetry. For example, if $g=\kappa_z$,
 \be
 \int\frac{d\omega d^2\kappa_\perp}{2(2\pi)^3}\frac{\hat{\varphi}_p(\omega,\kappa^2_\perp)}{\sqrt{(\omega+m_2)^2+\kappa^2_\perp}} \kappa_z=0,\label{A6}
 \en
because $\kappa_z$ is an odd function. The second case is $g=\kappa^2_z-\kappa^2_\perp/2$,
 \be
 \int\frac{d\omega d^2\kappa_\perp}{2(2\pi)^3}\frac{\hat{\varphi}_p(\omega,\kappa^2_\perp)}{\sqrt{(\omega+m_2)^2+\kappa^2_\perp}} \Bigg(\kappa^2_z-\frac{\kappa^2_\perp}{2}\Bigg)=0,\label{A7}
 \en
because the contributions of $\kappa^2_z$, $\kappa^2_{\perp1}$, and $\kappa^2_{\perp2}$ are equal.
The third case is $g=2 \kappa_z (e_2+\kappa_z)=\omega^2-m^2_2-\kappa^2_\perp$,
 \be
 \int\frac{d\omega d^2\kappa_\perp}{2(2\pi)^3}\frac{\hat{\varphi}_p(\omega,\kappa^2_\perp)}{\sqrt{(\omega+m_2)^2+\kappa^2_\perp}} (\omega^2-m^2_2-\kappa^2_\perp)=
 \int\frac{d\omega d^2\kappa_\perp}{2(2\pi)^3}\frac{\hat{\varphi}_p(\omega,\kappa^2_\perp)}{\sqrt{(\omega+m_2)^2+\kappa^2_\perp}} \kappa^2_\perp,\label{A8}
 \en
where Eqs. (\ref{A6}) and (\ref{A7}) are applied. We can employ Eqs. (\ref{A6}), (\ref{A7}) and (\ref{A8}) to prove that the integrations of $\omega$ for Eqs. (\ref{A32}), (\ref{A32p}), (\ref{T32}), and (\ref{T32p}) are all equal to zero.

\end{document}